\documentclass[11pt,a4paper]{article}
\usepackage{diagbox}
\usepackage{jheppub}
\usepackage{graphicx}
\usepackage{amsmath}
\usepackage{amssymb}
\usepackage{amsthm}
\usepackage{listings}
\usepackage{mathtools}
\usepackage{appendix}
\usepackage{mathrsfs}
\input epsf
\usepackage{epsfig}
\usepackage{tikz}
\usepackage{subfigure,tikz}
\usepackage{graphics}
\usepackage{graphicx}
\usepackage{diagbox}
\usepackage{mathtools}
\usepackage{xcolor}
\usepackage[thicklines]{cancel}
\usepackage{lscape}

\newcommand{\bea}{\begin{eqnarray}}
	\newcommand{\eea}{\end{eqnarray}}
\newcommand{\bean}{\begin{eqnarray*}}
	\newcommand{\eean}{\end{eqnarray*}}
\newcommand{\nn}{\nonumber \\}

\def\newline{{\hspace{15pt}}}

\def\eref#1{(\ref{#1})}

\def\d{{\rm d}}
\def\wt{\widetilde}

\def\d{\partial}
\def\dd{\text{d}}

\def\what{\widehat}
\newcommand{\overtr}[1]{(\overline{#1})}

\newcommand{\measure}[2]{{\langle{#1}\mathrm{d}^{#2}{#1}\rangle}}

\newcommand{\lt}[2]{\text{Lt}_{#1}^{(#2)}}

\def\co{\,,}
\def\ed{\,.}

\newcommand{\parall}[2]{{#1}\ /\kern -0.8em / \  {#2}}

\title{Solving arbitrary one-loop reduction via generating function}

\author[a,b]{Tingfei Li,}
\author[a]{Yuekai Song,}
\author[a,1]{Liang Zhang \note{Corresponding author.}}

\affiliation[a]{Zhejiang Institute of Modern Physics, Zhejiang University, Hangzhou, 310027, P. R. China }
\affiliation[b]{Kavli Institute for Theoretical Sciences (KITS), University of Chinese Academy of Sciences, Beijing 100190, China}

\emailAdd{tfli@zju.edu.cn, songyk@zju.edu.cn, liangzh@zju.edu.cn}

\abstract{Recently, the concept of generating function has been employed in one-loop reduction. For one-loop integrals encompassing arbitrary tensor ranks and higher-pole contributions, the generating function can be decomposed into a tensor part and a higher-pole part. While the tensor component has been thoroughly addressed in recent studies, there remains a lack of satisfactory investigations regarding the higher-pole part. In this work, we completely solve the problem. We first establish the partial differential equations governing the higher-pole generating function. Based on these equations, we derive an integration recursion relation and solve it iteratively. This approach enables us to explore the analytical structure of higher-pole reduction and provides a valuable tool for generating reduction coefficients efficiently.}


\begin{document}
	\maketitle
	\flushbottom

\section{Introduction} 

Since being studied extensively for over sixty years~\cite{Brown:1952eu,Melrose:1965kb}, the calculation of one-loop Feynman integrals does not pose any additional challenges. It is widely recognized that any one-loop integral can be expressed as a linear combination of master integrals (MIs)~
\cite{Stuart:1987tt,Devaraj_1998,vanOldenborgh:1989wn,Bern:1992em,DAVYDYCHEV1991107,Campbell:1996zw,Binoth:1999sp,Denner:2002ii,Denner:2005nn,Fleischer:2010sq,Passarino:1978jh,tHooft:1978jhc,vanNeerven:1983vr,Bern:1993kr,Bern:1994cg,Fleischer:1999hq,Duplancic:2003tv,Ellis:2007qk,Ossola:2006us}. Given that these MIs are already known, computing one-loop Feynman diagrams essentially becomes a reduction problem. Besides the traditional Passarino-Veltman (PV) reduction~\cite{Passarino:1978jh} which aims to solve the one-loop reduction, there are various methods available to one-loop or higher-loop reduction at the integrand and integral levels, such as Integration-By-Parts (IBP)~\cite{Chetyrkin:1981qh,Tkachov:1981wb,Laporta:2000dsw,vonManteuffel:2012np,vonManteuffel:2014ixa}, Ossola-Papadopoulos-Pittau(OPP) reduction~\cite{Ossola:2006us,Ossola:2007bb,Ellis:2007br}, Unitarity cut~\cite{Bern:1994zx,Bern:1994cg,Bern:1997sc,Britto:2004nc,Britto:2005ha,Anastasiou:2006gt,
    Britto:2006sj,Anastasiou:2006jv,Britto:2006fc,Britto:2007tt,Britto:2010um}, Intersection number \cite{Mastrolia:2018uzb,Mizera:2019ose,Frellesvig:2019uqt,Frellesvig:2019kgj,Mizera:2019vvs,Frellesvig:2020qot,Caron-Huot:2021xqj} and etc. Although these existing methods are valuable, it is still beneficial to explore the problem from new perspectives.

 Some recent papers~\cite{Feng:2021enk, Hu:2021nia, Feng:2022rwj,Feng:2022iuc,Feng:2022rfz,Li:2022cbx} introduced a new framework for general one-loop reduction. This framework utilizes an auxiliary vector, denoted as $R$, to represent a broad range of one-loop integrals that include both tensor structures and higher-poles:
\begin{align}\label{eq:one-loop}
    I_{n,\{v\}}^{(r)}\equiv  \int {d^D \ell \over i\pi^{D/2} } {(2R\cdot \ell)^r\over \prod_{i=1}^nD_i^{v_i+1}}\ed
\end{align}
Here, $D_i\equiv  (\ell-q_i)^2-m_i^2$ is the $i$-th inverse propagator and $v_i\ge 0$. The introduction of the auxiliary vector $R$ enables the solution of tensor reduction using differential operators of $R$, denoted as $\mathcal{D}_i = q_{i+1}\cdot \partial_R$, where $i=1,\ldots,n-1$, and $\mathcal{T}=\partial_R\cdot \partial_R$. It is worth mentioning that in the original articles~\cite{Feng:2021enk, Hu:2021nia}, $K_i$ was used instead of $q_{i+1}$. For convenience, we have set $q_1=0$. In this approach, the reduction coefficient needs to be expanded according to different tensor structures, the expansion coefficients being unknown. While the core idea of this method does not fundamentally deviate from the traditional Passarino-Veltman (PV) reduction~\cite{Passarino:1978jh}, its advantage lies in demonstrating that tensor reduction can be completely solved through recursion relations involving these expansion coefficients.

A recent alternative approach has been proposed in \cite{Feng:2022hyg} for solving tensor reduction of one-loop integrals, building upon previous works \cite{Feng:2021enk, Hu:2021nia}. This new approach offers a departure from previous methods by avoiding the need to explicitly consider the expansion of the reduction result for all possible tensor structures. Instead, it introduces a generating function that incorporates all relevant information, this idea can be traced back to~\cite{Ablinger:2014yaa,Kosower:2018obg} and followed by~\cite{Guan:2023avw,Hu:2023mgc,Feng:2024qsa}.The generating function depends on an auxiliary parameter $t$ and Lorentz scalars $R\cdot q_i$. By applying the same set of previously mentioned differential operators, a system of partial differential equations (PDEs) is derived. In principle, these PDEs can be solved by imposing appropriate boundary conditions. Notably, there exists a clear permutation symmetry among the equations derived from $\mathcal{D}_i$, which has the potential to simplify the solving process. However, the introduction of the condition $q_1=0$ breaks the permutation symmetry of the propagators. As a result, the process of finding solutions becomes significantly more challenging, especially for integrals involving a larger number of propagators.


 In contrast, the study of reduction in projective space offers a methodology that is not dependent on the number of propagators. The permutation symmetry among all propagators becomes transparent within the framework of projective space. Remarkably, it is possible to derive a straightforward recursion relation for one-loop tensor reduction, effectively transforming the reduction process into a purely algebraic computation \cite{Li:2022cbx}. Moreover, this recursion relation can be translated into an ordinary differential equation for the generating function. Solving this differential equation leads to a comprehensive analytical expression expressed in terms of hypergeometric functions.
 

 This example serves as a demonstration of the effectiveness of the generating function concept. The objective of this paper is to illustrate the applicability of generating functions in handling higher-pole cases in one-loop integrals. Unlike the tensor case, where a single parameter suffices to describe the integral, higher-pole integrals require the use of $n$ integers to fully characterize them. As a result, we introduce $n$ additional parameters in the generating function. Encouragingly, the generating function for higher-pole integrals exhibits a relatively simple form, representing integrals with mass-shifting. Building upon the findings in \cite{Feng:2022rwj}, which specifically addresses integrals with one higher pole (\emph{i.e.}, total power of $n+1$), we establish a set of partial differential equations. Initially, the task of solving the generating function may appear formidable.
 

Through a systematic elimination of the dependence on the mass-shifting variables one by one, the problem is eventually reduced to a straightforward ordinary differential equation, where the separation of variables becomes trivial. Consequently, the solution can be expressed as products of iterative integrals. Alternatively, the PDEs can be transformed into a simple recursion relation for integration, allowing any sector of the generating function to be solved as iterative integrals.\footnote{Similar formula also appears in \cite{Feng:2024qsa}.} By leveraging the properties of hypergeometric functions, it may be possible to obtain an analytical solution for these iterative integrals. Specifically, the $n\to n$ sector exhibits a concise expression, while the $n\to n-1$ sector can be expressed as a sum of hypergeometric functions. It is believed that the solution can be further simplified using hypergeometric functions to achieve a more elegant expression. For practical purposes, the integrand can be expanded as a Taylor series, making the integration trivial and resulting in the expression in terms of series summations. With the generating functions available for both the tensor case and the higher-pole case, obtaining the most general generating function becomes a straightforward task.
This paper covers the following: Section \ref{sec: review}: Brief review of reduction techniques for one-loop integrals using differential operators and projective space.
    Section \ref{sec:hp-gen}: Discussion of the solution for the higher-pole generating function.
    Section \ref{sec:example}: Illustration of our method through abstract examples with general $n$, along with specific calculations for bubble and triangle cases.
    Section \ref{discussion}: Discussions and future exploration.


\textbf{Notation and convention:}  In the following we will meet many symbols from methods in projective space. In section \ref{sec: review}, we introduce a $n \times n$ matrix $Q$ whose entries are $(Q)_{ij}=(m_i^2+m_j^2-(q_i-q_j)^2)/2$. $V_{(\boldsymbol{a};\boldsymbol{b})}$ is the submatrix of $V$ whose $\boldsymbol{a}$-th rows and $\boldsymbol{b}$-th columns have been removed.  If $\boldsymbol{a}=\boldsymbol{b}$ or $V$ is a vector(matrix in general sense), we simply denote as $V_{(\boldsymbol{a})}$. And we denote the matrix multiplication as $(\overline{AB})\equiv  AQ^{-1}B$, $(\widetilde{AB})\equiv  AQ^{*}B$, $(\overline{AB})_{(\boldsymbol{a};\boldsymbol{b})}\equiv  A_{(\boldsymbol{b})}(Q_{(\boldsymbol{a};\boldsymbol{b})})^{-1}B_{(\boldsymbol{a})}$, and $(\widetilde{AB})_{(\boldsymbol{a};\boldsymbol{b})}\equiv  A_{(\boldsymbol{b})}(Q_{(\boldsymbol{a};\boldsymbol{b})})^{*}B_{(\boldsymbol{a})}$, where $A$ and $B$ are vectors of length $n$, $Q^*,(Q_{(\boldsymbol{a};\boldsymbol{b})})^{*}$ are the adjugates of $Q,Q_{(\boldsymbol{a};\boldsymbol{b})}$.  In section \ref{sec:hp-gen} and after that, we will introduce the $Q$ matrix with mass shifting, \emph{i.e.}, $(Q)_{ij}=(m_i^2+m_j^2-(q_i-q_j)^2+u_i+u_j)/2$, and the previous $Q$ matrix without mass shifting will be denoted as $Q_0$. And we use a superscript $(0)$ in the denotation of matrix multiplication to indicate that the omitted $Q$ matrix is $Q_0$, such as  $(\overline{AB})^{(0)}\equiv  AQ_0^{-1}B$, and $(\overline{AB})^{(0)}_{(\boldsymbol{a};\boldsymbol{b})}\equiv  A_{(\boldsymbol{b})}(Q_{0(\boldsymbol{a};\boldsymbol{b})})^{-1}B_{(\boldsymbol{a})}$. The determinant of a matrix $A$ is denoted as $|A|$.

	
\section{Review of one-loop integral reduction in projective space}
\label{sec: review}
As mentioned in the introduction, our target is to reduce a general one-loop integral encompassing both tensor structures and higher-pole contributions 
\begin{align}\label{eq:general-generating-func}
    I_{n,\{v\}}^{(r)}\equiv  \int {d^D \ell \over i\pi^{D/2} } {(2R\cdot \ell)^r\over \prod_{i=1}^nD_i^{v_i+1}}\co
\end{align}
with $D_i\equiv  (\ell-q_i)^2-m_i^2$ being the $i$-th inverse propagator. One can define a generating function for them 
\begin{align}
   	     G_n(t,\{u\})&\equiv \sum_{r=0}^{\infty}\sum_{\{v\}=0}^{\infty}\left(\prod_{i=1}^n u_i^{v_i} \right)t^r \int {d^D \ell \over i\pi^{D/2} } {(2R\cdot \ell)^r\over \prod_{i=1}^n D_i^{v_i+1}} \nn
   	     &=\int {d^D \ell \over i\pi^{D/2} } {1\over 1-t(2R\cdot \ell)}{1\over \prod_{i=1}^n(D_i-u_i)} \ed
\end{align}
Since the reduction of $I_{n,{v}}^{(r)}$ to master integrals is well-defined, \emph{i.e.}
\begin{align}
I_{n,\{v\}}^{(r)}=\sum_{\boldsymbol{b}}C^{(r)}_{n,\{v\}\to \widehat{\boldsymbol{b}}}I_{n;\widehat{\boldsymbol{b}}}\co
\end{align}
so is the reduction of $G_n(t,{u})$
\begin{align}
G_n(t,\{u\})=\sum_{\boldsymbol{b}}\mathcal{G}_{n\to \widehat{\boldsymbol{b}}}(t,\{u\})I_{n;\widehat{\boldsymbol{b}}}\co
\end{align}
where\footnote{The convergence of this series is guaranteed since both $t$ and $\{u\}$ are small quantities.}
\begin{align}
\mathcal{G}_{n\to \widehat{\boldsymbol{b}}}(t,\{u\})=\sum_{r=0}^{\infty}\sum_{\{v\}=0}^{\infty}\left[\left(\prod_{i=1}^n u_i^{v_i} \right)t^r C^{(r)}_{n,\{v\}\to \widehat{\boldsymbol{b}}}\right]\ed
\end{align}
 
 One can see the generating function for the combined case is just the pure tensor case with mass-shifting. So one can do the tensor reduction first, then deal with the scalar integrals with mass-shifting:
\begin{align}
   	    G_n(t,\{u\})=\sum_{\boldsymbol{a}} G_{n\to \widehat{\boldsymbol{a}}}(t,\{u\})F_{n;\widehat{\boldsymbol{a}}}(\{u\})\co
\end{align}
where $F_{n;\widehat{\boldsymbol{a}}}(\{u\})$ is the scalar integrals with shifting mass $m_i'^2=m_i^2+u_i$ and $\widehat{\boldsymbol{a}}$ means to remove the propagators in the label set $\boldsymbol{a}$, and  $F_{n;\widehat{\boldsymbol{a}}}$ is the function of the variables $\{u\backslash\boldsymbol{u}_a\}$, a subset of $\{u\}$ without any element corresponding to propagators in $\boldsymbol{a}$,  but we can also see it as a function of $\{u\}$ in general sense. And for later use, for $f(\{u\})$ which is a general function of $\{u\}$, we define
\begin{align}
\label{def_reduce}
    f(\{u\backslash\boldsymbol{u}_a\})\equiv f(\{u\})\vert_{u_j=0,\forall j \in \boldsymbol{a}}\co
\end{align}
To avoid any confusion, we declare in this paper, if the dependence on the variables of a general function of $\{u\}$ is not explicitly written, it always means the variables are $\{u\}$ (maybe in general sense). And $F_n$ can be also regarded as the generating function of higher-pole case, defined as 
\begin{align}
    F_n(\{u\})\equiv  \int {d^D \ell \over i\pi^{D/2} } {1\over \prod_{i=1}^n(D_i-u_i)}=\sum_{\{v\}=0}^{\infty}  \int {d^D \ell \over i\pi^{D/2} } \prod_{i=1}^n {u_i^{v_i} \over  D_i^{v_i+1}}\ed
\end{align}
The expression of $G_{n\to n;\widehat{\boldsymbol{a}}}(t,\{u\})$ is the same as that in \cite{Hu:2023mgc} but with mass-shifting. So, the remaining problem is to handle the generating function for the higher-pole case. Here we briefly review the one-loop reduction in projective space, for it gives a simple recursion relation which makes it possible to find an explicit expression of one-loop tensor generating function. 

As pointed out in \cite{Arkani-Hamed:2017ahv,Feng:2022rwj}, one can write the general one-loop integral in projective space with a compact form. To simplify our denotation, we first introduce the integrals in projective  space
	\begin{align}
		E_{n,k}[T]\equiv  \int_{\Delta}{\measure{X}{n-1}T[X^k]\over (XQX)^{n+k\over 2}}\ ;~~ T[X^k]\equiv  T_{I_1I_2\ldots I_k}X^{I_1}X^{I_2}\cdots X^{I_k}\co
	\end{align}
	where  $T$ is a general rank-$k$ tensor and $\Delta$ is a simplex in $n$-dimensional space defined by $X_I>0,\forall I=1,2,\ldots,n$.  The homogeneous coordinates $X_I$ are denoted by a square bracket $X=[x_1:x_2:\ldots:x_n]$, and two coordinates are equivalent to each other up to a scaling, \textit{\textit{i.e}.}, $[x_1:x_2:\ldots:x_n]\sim [kx_1:kx_2:\ldots:kx_n]$ for any $k\neq 0$. The measure in the projective space is given by the differential form
	\begin{align}
		\measure{X}{n-1}={\epsilon^{I_1,I_2,\ldots,I_n}\over (n-1)!}X_{I_1} \dd X_{I_2}\wedge \dd X_{I_3} \wedge \ldots \wedge \dd X_{I_n}\ed
	\end{align}
	The matrix $Q$ appearing in the denominator $
	XQX=Q^{IJ}X_IX_J $ is defined previously. 
	For simplicity, we denote $E_{n,k}[V^i]\equiv  E_{n,k}[T=\otimes V^i\otimes L^{k-i}]$ for arbitrary vector $V$ and the constant vector $L\equiv  [1:1:\cdots:1]$:
	\begin{align}
		\otimes V^i\otimes L^{k-i}\equiv  \underbrace{V\otimes V\otimes \dots V}_{i\ \text{times}} \otimes \underbrace{L\otimes L\otimes \dots L}_{k-i\ \text{times}}.
	\end{align}
	The order of $V$ and $L$ is irrelevant since both are contracted with $X$'s. After some algebra \cite{Feng:2022rwj},
	we find \eref{eq:general-generating-func} can be written as a compact form in terms of $E_{n,k}$
	\begin{align}
	I_{n,\{v\}}^{(r)}=&\sum_{i=0}^{r} {i!\Gamma(\bar{v}-D/2-r)\over (-1)^{\bar{v}+r}  (\bar{v}-n+i)!}\mathscr{C}^{D/2+r-\bar{v}}_{r,i}(R^2)^{r-i\over 2} E_{n,2\bar{v}-n-D-r+i}[ S^{\bar{v}-n+i}]\Big\vert_{t^i\mathbf{z}^{\mathbf{v}-1}}\co~~~~~~~\label{UniversalCoeff}
	\end{align}
	where $\bar{v}\equiv  \sum_{i=1}^n v_i,S\equiv  tV+Z,Z\equiv \sum_{i=1}^{n}z_iH_i$ and $\vert_{t^i\mathbf{z}^{\mathbf{v}-1}}$ means to take the coefficient of $t^i\mathbf{z}^{\mathbf{v}-1}\equiv  t^i\prod_{i=1}^n z_i^{v_i-1}$ and where we have divide the symmetry factor ${\mathsf{S}[i,z^{\mathbf{v}_n}]}=\binom{\bar{v}-n+i}{i}\binom{\bar{v}-n}{v_1-1}\binom{\bar{v}-n-v_1}{v_2-1}\ldots \binom{v_n-1}{v_n-1}$. We have defined the vectors $V$, $H_i$ as 
	\begin{align}
	V\equiv  [R\cdot q_1: R\cdot q_2:\ldots: R\cdot q_n],~~~~ H_i\equiv  [0: \ldots 0:\mathop{1}\limits_{{\text{i-th}}}:0:\ldots: 0]\co
	\end{align}
	and the expansion coefficient in \eref{UniversalCoeff} is
	\begin{align}
	\mathscr{C}^k_{r,i}=\frac{2^r r! \Gamma\big(  {r-i+1\over 2}\big)}{\sqrt{\pi }
		i! (r-i)!}\prod_{j=1}^{{r+i\over 2}}(k+1-j)=\frac{2^r r!k! \Gamma\big(  {r-i+1\over 2}\big)}{\sqrt{\pi }
		i! (r-i)!(k-{r+i\over 2})!}\co~~~ {r-i\over 2}\in \mathbb{N},
	\end{align}
	where we require $i$ to have the same parity as $r$.  We denote $I_{v;D}^{(r)}$ as the one-loop integrals It is found that the reduction of general one-loop integrals is solved by a simple recursion relation for $E_{n,k}[T]$
	\begin{align}
	E_{n,k}[(Q\widetilde{Q}T)]=\alpha_{n,k}\left[\sum_{b=1}^nE^{(b)}_{n-1,k-1}[(H_b\widetilde{Q}T)]+\sum_{k-1\ \text{ways}}E_{n,k-2}[\text{tr}_{\widetilde{Q}} T]\right]\co~~~~\label{general-resur}
	\end{align}
	where  $\alpha_{n,k}={1\over n+k-2}$ and $\wt{Q}$ is an arbitrary symmetric matrix. In the formula we need to sum the $(k-1)$ ways to contract indices between $\wt{Q}$ and $T$
	\begin{align}
		(Q\widetilde{Q}T)^{I_1I_2,\ldots,I_k}&=Q_0^{I_1J_1}\widetilde{Q}_{J_1J_2}T^{J_2,I_2,I_3,\ldots,I_k}\co\nn
		\sum_{(k-1) \text{ways}}(\text{tr}_{\widetilde{Q}} T)^{I_3,\ldots,I_k}&=\widetilde{Q}_{I_1I_2}T^{I_1I_2I_3\ldots I_k}+\widetilde{Q}_{I_1I_2}T^{I_1I_3I_2 \ldots I_k}+\cdots +\widetilde{Q}_{I_1I_2}T^{I_1I_3\ldots I_{k}I_2}\ed
	\end{align}
	For non-degenerate $Q$, we just take $\widetilde{Q}=Q^{-1}$, after some algebra one can find there exists a non-trivial recursion relation for one-loop tensor integrals \cite{Li:2022cbx}
		\begin{align}
				I_{n}^{(r)}={1\over (\overline{LL})}\left[A_rI_{n}^{(r-1)}+B_rI_{n}^{(r-2)}+\lt{n}{r}\right],
			\label{eq:relation 01}
		\end{align}
	where the coefficients are
		\begin{align} \label{eq:define XY}
				A_r=&{2(D+2r-n-2)\over D+r-n-1}(\overline{VL})\co \nn
		B_r=&{4(r-1)(R^2-(\overline{VV}))\over  D+r-n-1}\co \nn
  	\lt{n}{r}=&\left[(\overline{H_bL})(\overline{VL})_{(b)}-(\overline{H_bV})(\overline{LL})_{(b)}\right]I^{(r-1)}_{n;\what{b}}\nn
		&+{2(r-1)\left[(\overline{H_bL})R^2+(\overline{H_bV})(\overline{VL})_{(b)}-(\overline{H_bL})(\overline{VV})_{(b)}\right]\over D+r-n-1 }I_{n;\what{b}}^{(r-2)}\nn
		\equiv & A_{r;\what{b}}I_{n;\what{b}}^{(r-1)}+B_{r;\what{b}}I_{n;\what{b}}^{(r-2)}\ed
		\end{align}
We multiply both sides of equation \eqref{eq:relation 01} by $t^r$ and sum over $r$ from $0$ to $\infty$, the LHS is the tensor generating function
	\begin{equation}
			G_n({t})=\sum_{r=0}^{\infty}{t}^r\cdot I^{(r)}_n=\int\frac{d^Dl}{i\pi^{D/2}}\frac{1}{1-{t}(2R\cdot l)}\frac{1}{\prod^n_{j=1}(l-q_j)^2-m_j^2}\ed
	\end{equation}
After some algebra, we can get a differential function\footnote{The details can be found in \cite{Hu:2023mgc}.}
	\begin{equation}
		\begin{aligned}
                &\bigg((D-n-1)-2(D-n)\cdot\frac{(\overline{VL})}{(\overline{LL})}t-4\cdot\frac{R^2-(\overline{VV})}{(\overline{LL})}t^2\bigg)G _n(t)\\
			+&\bigg(t-4\cdot\frac{(\overline{VL})}{(\overline{LL})}t^2-4\cdot\frac{R^2-(\overline{VV})}{(\overline{LL})}t^3\bigg)G'_n(t)-(D-n-1)I_n\\
			=&\sum_{b=1}^n \Bigg\{X^{(b)}\bigg((D-n)t\cdot G_{n;\widehat{b}}(t)+t^2\cdot G'_{n;\widehat{b}}(t)\bigg)+2Y^{(b)}\bigg(t^3\cdot G'_{n;\widehat{b}}(t)+t^2\cdot G_{n;\widehat{b}}(t)\bigg)\Bigg\},
		\end{aligned}
		\label{eq:relation 001}
	\end{equation}
 where
 \begin{align}
			X^{(b)}=&\left((\overline{H_bL})(\overline{VL})_{(b)}-(\overline{H_bV})(\overline{LL})_{(b)}\right)/(\overline{LL}),\\
			Y^{(b)}=&\left((\overline{H_bL})R^2+(\overline{H_bV})(\overline{VL})_{(b)}-(\overline{H_bL})(\overline{VV})_{(b)}\right)/(\overline{LL}).
		\end{align}
Then we expand the results to  the  master integrals,
\begin{equation}
\begin{aligned}
&I_{n}^{(r)}=\sum_{\boldsymbol{b}} C^{(r)}_{n\to\widehat{\boldsymbol{b}}}~    I_{n;\widehat{\boldsymbol{b}}}\ ,\\ 
&G_n(t)=\sum_{\boldsymbol{b}} \Big\{\sum_{r=0}^\infty t^r\cdot C^{(r)}_{n\to \widehat{\boldsymbol{b}}}\Big\}~ I_{n;\widehat{\boldsymbol{b}}}=\sum_{\boldsymbol{b}} G_{n\to \widehat{\boldsymbol{b}}}(t) I_{n,\widehat{\boldsymbol{b}}}\ .
\end{aligned}
\end{equation}
Finally the equation  \eqref{eq:relation 001} is transformed into a recursive formula of the reduction coefficients,
\begin{equation}
    \begin{aligned}
        &\bigg((D-n-1)-2(D-n)\cdot\frac{(\overline{VL})}{(\overline{LL})}t-4\cdot\frac{R^2-(\overline{VV})}{(\overline{LL})}t^2\bigg)G_{n\to \widehat{\boldsymbol{b}}}(t)\\
			&+\bigg(t-4\cdot\frac{(\overline{VL})}{(\overline{LL})}t^2-4\cdot\frac{R^2-(\overline{VV})}{(\overline{LL})}t^3\bigg)G'_{n\to \widehat{\boldsymbol{b}}}(t)\\
			=&\sum_{b_i\in \boldsymbol{b}} \Bigg\{X^{(b_i)}\bigg((D-n)t\cdot G_{n,\widehat{b_i}\to \widehat{\boldsymbol{b}}}(t)+t^2\cdot  G'_{n,\widehat{b_i}\to \widehat{\boldsymbol{b}}}(t)\bigg)\\
   &+2Y^{(b_i)}\bigg(t^3\cdot  G'_{n,\widehat{b_i}\to \widehat{\boldsymbol{b}}}(t)+t^2\cdot G_{n,\widehat{b_i}\to \widehat{\boldsymbol{b}}}(t)\bigg)\Bigg\}+(D-n-1)C^{(0)}_{n\to \widehat{\boldsymbol{b}}}\ed
    \end{aligned}
    \label{eq:relation 02}
\end{equation}
In this paper, we use subscript $(n,\widehat{\boldsymbol{a}}\to \widehat{\boldsymbol{b}})$ to indicate the reduction of a $n$-point integrals with propagators in label set $\boldsymbol{a}$ removed to the one with propagators in $\boldsymbol{b}$ removed. By definition, the reduction coefficient  $C^{(0)}_{n\to \widehat{\boldsymbol{b}}}=1$ only when $ \boldsymbol{b}=\emptyset$ (\emph{i.e.}, no propagators removed). Otherwise $C^{(0)}_{n\to \widehat{\boldsymbol{b}}}=0$. One can solve the differential equation recursively.  Initially, by setting $\boldsymbol{b}=\emptyset$ in \eref{eq:relation 02} makes the terms within the curly braces on the right-hand side vanish. This leads to a first-order non-homogeneous ordinary differential equation for $G_{n\to n}({t})$, which can be readily solved using the initial condition. Subsequently, by setting $\boldsymbol{b}=\{b_1\}$ (\emph{i.e.}, removing one propagator), we can solve the differential equation for $n$-gon to $(n-1)$-gon sector. Repeating this process allows us to obtain all reduction coefficients for the pure tensor generating functions recursively.

\section{Higher-pole generating function}
\label{sec:hp-gen}
As discussed in the previous section, we now turn to deal with the higher-pole generating function. We first establish an integration recursion relation, then derive the explicit expressions for $n\to n-1$ and $n\to n-2$ sectors. Finally we apply the recursion relation to obtain a general expression for the $n\to n-k$ sector.
\subsection{Recursion relation}
We can expand the higher-pole generating function to the master basis as below
  \begin{align}
    F_n(\{u\})=\sum_{\boldsymbol{a}} \mathcal{F}_{n\to\widehat{\boldsymbol{a}}}(\{u\})I_{n;\widehat{\boldsymbol{a}}} \ed
  \end{align}
  Similar to the tensor case, $n$ differential equations are expected to determine $F_n$'s expression. So we consider taking derivation over the $i$-th parameter $u_i$, then we have 
  \begin{align}
      {\partial\over \partial u_i}F_{n}(\{u\})=\int {d^D \ell \over i\pi^{D/2} } {1\over \prod_{j=1}^n(D_j-u_j)^{1+\delta_{ij}}}\ed
  \end{align}

It can be regarded as a one-loop integral with mass shifting and only one propagator of power 2. And it is easy to reduce it to the basis without higher poles. 
We can just use the results in \cite{Feng:2022rwj} for scalar pentagon with only one higher pole, which is easy to generalize to arbitrary $n$-gon integral:
	\begin{align}\label{eq: Ccoefficient}
	C^{(Z)}_{n \to \widehat{ijkl}}&=\frac{1}{32} (D-n+3) \overtr{L L}_{(ijkl)} \Big(\overtr{H_i L}_{(jk)} \overtr{H_j Z} \overtr{H_k L}_{(j)} \overtr{H_l L}_{(ijk)}+\text{Perm}(ijkl)\Big)\co\nn
	C^{(Z)}_{n\to \widehat{ijk}}&=-\frac{1}{16} (D-n+2) \overtr{L L}_{(ijk)} \Big(\overtr{H_i L}_{(j)} \overtr{H_k L}_{(ij)} \overtr{H_j Z}+\text{Perm}(ijk)\Big)\co\nn
	C^{(Z)}_{ n\to \widehat{ij}}&=\frac{1}{8} (D-n+1) \overtr{L L}_{(ij)} \left(\overtr{H_j L}_{(i)} \overtr{H_i Z}+(i\leftrightarrow j)\right)\co\nn
C^{(Z)}_{n\to \widehat{i}}&=-\frac{1}{4} (D-n) \overtr{L L}_{(i)} \overtr{H_i Z}\co\nn
	C^{(Z)}_{n\to \widehat{\emptyset}}&=\frac{1}{2} (D-n-1) \overtr{Z L}\ed 
	\end{align}
 Here ``Perm" means all the permutations of the deleted labels, while $Z$ is an arbitrary auxiliary vector $Z=[z_1,z_2,\ldots,z_n]=\sum_i z_i H_i$, 
 and $C^{(Z)}_{n\to\widehat{\boldsymbol{a}}}$ corresponds to the reduction of $\sum_i z_i\partial_{u_i} F_n $, \emph{i.e.},
    \begin{align}
        \sum_i z_i\partial_{u_i} F_n(\{u\}) = \sum_{\boldsymbol{a}}C^{(Z)}_{n\to\widehat{\boldsymbol{a}}} F_{n,\widehat{\boldsymbol{a}}}(\{u\})\ed
    \end{align}
    Taking a given $z_i$ to $1$ and others to 0, we get $n$ partial differential equations
    \begin{align}
        \partial_{u_i} F_n(\{u\}) = \sum_{\boldsymbol{a}}C_{n\to\widehat{\boldsymbol{a}}}^{(H_i)}F_{n,\widehat{\boldsymbol{a}}}(\{u\}), ~~~1\leq i\leq n \ed\label{pdes}
    \end{align}
    We could consider dropping the mass-shifting $\{u\}$ one by one to solve the equations. First, we treat $u_i$ as the only variable with all the other $u_j$ becoming part of new masses ${m'_j}^{2}=m_j^2+u_j$, and regard $F_{n;\widehat{\boldsymbol{a}}}(\{u \backslash u_i\})$ (see definition \eref{def_reduce}) as the master integrals, so we have the decomposition
    \begin{align}
        F_{n}(\{u\})\equiv\sum_{\boldsymbol{a}}\mathcal{F}_{n\to\widehat{\boldsymbol{a}}}^{(u_{i})}(\{u\})F_{n,\widehat{\boldsymbol{a}}}(\{u\backslash u_i\}) \ed\label{drop_u1}
  \end{align}
Applying this formula for $u_1$, we obtain a modified generating function with the $u_1$ dependence eliminated. We can then iteratively remove the remaining $u_j$ dependences using the same strategy, ultimately resulting in
    \begin{align}
    \label{chain}
    F_{n}(\{u\})=\sum_{\boldsymbol{b}_1\subseteq \ldots\subseteq \boldsymbol{b}_n}\mathcal{F}_{n\to\widehat{\boldsymbol{b}}_1}^{(u_{1})}(\{u\})\mathcal{F}_{n,\widehat{\boldsymbol{b}}_1\to\widehat{\boldsymbol{b}}_2}^{(u_{2})}(\{u\backslash u_1\})\cdots \mathcal{F}_{n,\widehat{\boldsymbol{b}}_{n-1}\to\widehat{\boldsymbol{b}}_n}^{(u_{n})}(\{u\backslash\boldsymbol{u}_{n-1}\})I_{n;\widehat{\boldsymbol{b}}_n}\co
    \end{align}
    where $\boldsymbol{u}_{i}$ denotes the set $\{u_1,u_2,\ldots,u_{i}\}$ and we just write the singleton set $\{u_i\}$ as $u_i$. We have defined the subscript $(n,\widehat{\boldsymbol{a}}\to \widehat{\boldsymbol{b}})$ after \eref{eq:relation 02}, which gives
    \begin{align}
          F_{n,\widehat{\boldsymbol{a}}}(\{u\})\equiv\sum_{\boldsymbol{b};\boldsymbol{a}\subseteq \boldsymbol{b}}\mathcal{F}_{n,\widehat{\boldsymbol{a}}\to\widehat{\boldsymbol{b}}}^{(u_{i})}(\{u\})F_{n,\widehat{\boldsymbol{b}}}(\{u\backslash u_i\})\co
    \end{align}
    and in \eref{chain} we have also used the definition \eref{def_reduce}
    \begin{align}
    \mathcal{F}_{n,\widehat{\boldsymbol{b}}_i\to\widehat{\boldsymbol{b}}_{i+1}}^{(u_{i+1})}(\{u\backslash \boldsymbol{u}_i\})\equiv \mathcal{F}_{n,\widehat{\boldsymbol{b}}_i\to\widehat{\boldsymbol{b}}_{i+1}}^{(u_{i+1})}(\{u\})\Big|_{\boldsymbol{u}_i=0}\ed
    \end{align}
   Note that we do not demand $\boldsymbol{u}_i$ corresponds to the propagators in label set $\boldsymbol{b}_i$, thus $\boldsymbol{u}_i$ in  $F_{n,\widehat{\boldsymbol{b}}_i}(\{u\})$ and $\mathcal{F}_{n,\widehat{\boldsymbol{b}}_i\to\widehat{\boldsymbol{b}}_{i+1}}^{(u_{i+1})}(\{u\})$ which are seen as general functions of $\{u\}$, may not be 0, so it is necessary to emphasize the dependence on variables in  $\mathcal{F}_{n,\widehat{\boldsymbol{b}}_i\to\widehat{\boldsymbol{b}}_{i+1}}^{(u_{i+1})}(\{u\backslash \boldsymbol{u}_i\})$ is necessary. Similar explanations also apply to functions like $F_{n,\widehat{\boldsymbol{b}}}(\{u\backslash u_i\})$.
    
    Then we try to reduce the whole problem of PDE to ODE\footnote{Treat other $u_{i\neq 1}$ as constants.} of ${\cal F}_{n\to \widehat{\boldsymbol{a}}}^{(u_1)}$. Using \eref{drop_u1} and \eref{pdes}, then we have
    \begin{align}\label{eq:ODE}  \sum_{\boldsymbol{a}}\left(\partial_{u_{1}}\mathcal{F}_{n\to\widehat{\boldsymbol{a}}}^{(u_{1})}\right)F_{n,\widehat{\boldsymbol{a}}}(\{u \backslash u_1\}) & =\sum_{\boldsymbol{b}}C_{n\to\widehat{\boldsymbol{b}}}^{(H_1)}F_{n,\widehat{\boldsymbol{b}}}(\{u\})\ed
    \end{align}
    Depending on whether $1$ belongs to $\boldsymbol{b}$, we split the term on the RHS in \eref{eq:ODE}
    \begin{align}
    \sum_{\boldsymbol{a}}\left(\partial_{u_{1}}\mathcal{F}_{n\to\widehat{\boldsymbol{a}}}^{(u_{1})}\right)F_{n,\widehat{\boldsymbol{a}}}(\{u \backslash u_1\})=&\sum_{\boldsymbol{b};1\in \boldsymbol{b}}C^{(H_1)}_{n\to\widehat{\boldsymbol{b}}}F_{n,\widehat{\boldsymbol{b}}}(\{u\backslash u_1\})\nn
    	& +\sum_{\boldsymbol{b};1\notin \boldsymbol{b}}C^{(H_1)}_{n\to\widehat{\boldsymbol{b}}}\sum_{\boldsymbol{c};\boldsymbol{b}\subseteq \boldsymbol{c}}\mathcal{F}_{n,\widehat{\boldsymbol{b}}\to\widehat{\boldsymbol{c}}}^{(u_{1})}(\{u\})F_{n,\widehat{\boldsymbol{c}}}(\{u \backslash u_1\})\ed
    \end{align}
    Regarding $F_{n;\widehat{\boldsymbol{a}}}[\{u\backslash u_1\}]$ as basis and comparing the both sides, we have
    \begin{align}\label{eq:ui-recur}
    	\partial_{u_{1}}\mathcal{F}_{n\to\widehat{\boldsymbol{a}}}^{(u_{1})} & =\delta(1\in \boldsymbol{a})C^{(H_1)}_{n\to\widehat{\boldsymbol{a}}}+\sum_{\boldsymbol{b};1\notin \boldsymbol{b}\subset \boldsymbol{a}}C^{(H_1)}_{n\to\widehat{\boldsymbol{b}}}\mathcal{F}_{n,\widehat{\boldsymbol{b}}\to\widehat{\boldsymbol{a}}}^{(u_{1})}(\{u\})\ed
    \end{align} 
    where we define a similar Keronecker function 
    \begin{align}
        \delta(1\in\boldsymbol{a})=\begin{cases}
        1 & \text{if}\ 1\in\boldsymbol{a} \co\\
        0 & \text{if}\ 1\notin\boldsymbol{a} \ed
        \end{cases}
    \end{align}
    We will show that the formula gives an integration recursion for $\mathcal{F}_{n\to\widehat{\boldsymbol{a}}}^{(u_{1})}$. 
    To see this, we first consider the simplest case $\boldsymbol{a}=\emptyset$. Then we have the ODE of $n\to n$
    \begin{align}
\partial_{u_{1}}\mathcal{F}_{n\to\widehat{\emptyset}}^{(u_{1})}=C_{n\to\widehat{\emptyset}}^{(H_1)}\mathcal{F}_{n\to\widehat{\emptyset}}^{(u_{1})}\ed
    \end{align}  
    Using the result in \eqref{eq: Ccoefficient}, we have
    \begin{align}
        \log \mathcal{F}_{n\to\widehat{\emptyset}}^{(u_{1})} =  \gamma_n \int  \frac{ \widetilde{H_iL}} {|Q|} d{u_1} \co
    \end{align}
    where we have defined 
    \begin{align}
        \gamma_n \equiv \frac{1}{2} (D-n-1)\ed
    \end{align}
    To perform the integration, we noticed 
    \begin{align}
        \d_{u_i} |Q| = \widetilde{H_iL}\ed
    \end{align}
    which can be checked using direct matrix calculation, thus
    \begin{align}
        \log \mathcal{F}_{n\to\widehat{\emptyset}}^{(u_{1})} =   \gamma_n \log |Q| +c\co
    \end{align}

    With the obvios boundary condition $\lim_{u_1\to 0}\mathcal{F}_{n\to\widehat{\emptyset}}^{(u_1)}=1$, we obtain a compact expression 
    \begin{align}\label{eq:n2nCoeff}
        \mathcal{F}_{n\to\widehat{\emptyset}}^{(u_1)}=\left(\frac{|Q|}{|Q|_{{u_1=0}}}\right)^{\gamma_n}\ed
    \end{align}
    Thus we can solve the  $n\to n$ section of the higher-pole generating function by \eref{chain} 
    \begin{align}\label{eq:ntoncoefficient}
        \mathcal{F}_{n\to\widehat{\emptyset}}=\left(\frac{|Q|}{|Q|_{{u_1=0}}}\right)^{\gamma_n}\left(\frac{|Q|_{{u_1=0}}}{|Q|_{{\boldsymbol{u_2}=0}}}\right)^{\gamma_n}\cdots \left(\frac{|Q|_{{\boldsymbol{u_{n-1}}=0}}}{|Q|_{{\boldsymbol{u_{n}}=0}}}\right)^{\gamma_n}=\left(\frac{|Q|}{|Q_0|}\right)^{\gamma_n}\ed
    \end{align}
   
    Then we try to solve the next sector $n\to n-1$.  We just give an easy example where  $\boldsymbol{a}=\{1\}$.\footnote{When $\boldsymbol{a}\neq\{1\}$, the situation may become slightly more complex. However, since we will later employ a more concise method to calculate $n\to n-1$, we do not delve into these intricate cases here.} Now \eref{eq:ui-recur} becomes
    \begin{align} \partial_{u_{1}}\mathcal{F}_{n\to\widehat{1}}^{(u_{1})}=C_{n\to\widehat{1}}^{(H_1)}+C_{n\to\widehat{\emptyset}}^{(H_1)}\mathcal{F}_{n\to\widehat{1}}^{(u_{1})}\ed
    \end{align}
    Substituting the \eqref{eq: Ccoefficient} into it, we get
    \begin{align}
    	\d_{u_1} \mathcal{F}_{n\to\widehat{1}}^{(u_{1})}=\frac{-1}{4}(D-n)\overtr{LL}_{(1)}\overtr{H_1H_1}+\gamma_n\overtr{H_1L}\mathcal{F}_{n\to\widehat{1}}^{(u_{1})}\ed
    \end{align}
Reorganizing it, we obtain
    \begin{align}
    	|Q|\d_{u_1}\mathcal{F}_{n\to\widehat{1}}^{(u_{1})}-\gamma_n ( \widetilde{H_iL})\mathcal{F}_{n\to\widehat{1}}^{(u_{1})}=\frac{-1}{4}(D-n)\overtr{LL}_{(1)}Q^*_{11}\ed
    \end{align}
    We can choose $F_{n\to\widehat{1}}^{(u_{1})}=|Q|^{\gamma_n}f$, then we have
    \begin{align}
    	\d_{u_1}f=|Q|^{-\gamma_n}\frac{-1}{4}(D-n)\overtr{LL}_{(1)}Q^*_{11}\ed
    \end{align}
    It can be solved by direct integration. With the boundary condition $\lim_{u_1\to0}\mathcal{F}_{n\to\widehat{1}}^{(u_{1})}=0$, we find the solution is
   \begin{align}
        \mathcal{F}_{n\to\widehat{1}}^{(u_{1})}=\frac{-1}{4}(D-n)\overtr{LL}_{(1)}Q^*_{11} \Big[\mathcal{G}(u_1)-\mathcal{G}(0)\Big]\co
    \end{align}
    where  
    \begin{align}
        \mathcal{G}(u_1)=\frac{\left(u_1-x_1\right)
   \left(\frac{u_1-x_2}{x_1-x_2}\right)
   {}^{-\gamma_n} \,
   _2F_1\left(-\gamma_n,\gamma_n+1;\gamma_n+2;\frac{x_1-u_1}
   {x_1-x_2}\right)}{\gamma_n+1}\co
    \end{align}
    Here, $x_1,x_2$ are the roots of $|Q|(u_1)=0$ (regarding $u_{i>1}$ as constant), with
   \begin{align}\label{expansion_q}
    |Q|&=|Q_{0}|+(\widetilde{H_iL})^{(0)}u_i+\frac{(-1)^{i+j+1}}{4}(\widetilde{LL})^{(0)}_{(i;j)}u_iu_j \nn
    &\equiv  |Q_0|+|Q|^{(k)}u_k+|Q|^{(kl)}u_ku_l\ed
    \end{align}
    Similarly, for the general case, we have
    \begin{align}
        \partial_{u_{1}}\mathcal{F}_{n\to\widehat{\boldsymbol{a}}}^{(u_{1})}-C_{n\to\emptyset}^{(H_1)}\mathcal{F}_{n\to\widehat{\boldsymbol{a}}}^{(u_{1})}=\delta(1\in \boldsymbol{a})C_{n\to\widehat{\boldsymbol{a}}}^{(H_1)}+\sum_{\boldsymbol{b}\neq \emptyset,1\notin \boldsymbol{b}\subseteq \boldsymbol{a}}C_{n\to\widehat{\boldsymbol{b}}}^{(H_1)}\mathcal{F}_{n,\widehat{\boldsymbol{b}}\to\widehat{\boldsymbol{a}}}^{(u_{1})}\ed
    \end{align}
    So we can always set 
    \begin{align}
        \mathcal{F}_{n\to\widehat{\boldsymbol{a}}}^{(u_{1})}=|Q|^{\gamma_n}f_{n\to\widehat{\boldsymbol{a}}}^{(u_{1})}\co
    \end{align}
    then we get
    \begin{align}\label{eq:uintrecur}
    \mathcal{F}_{n\to\widehat{\boldsymbol{a}}}^{(u_{1})}=\delta_{|\boldsymbol{a}|,0}\mathcal{F}^{(u_1)}_{n\to \widehat{\emptyset}}+|Q|^{\gamma_n}\int_{0}^{u_{1}}du_{1}|Q|^{-\gamma_n}\left(\delta(1\in \boldsymbol{a})C_{n\to\widehat{\boldsymbol{a}}}^{(H_1)}+\sum_{\boldsymbol{b}\not=\emptyset,u_{1}\notin \boldsymbol{b},\boldsymbol{b}\subseteq \boldsymbol{a}}C_{n\to\widehat{\boldsymbol{b}}}^{(H_1)}\mathcal{F}_{n,\widehat{\boldsymbol{b}}\to\widehat{\boldsymbol{a}}}^{(u_{1})}\right)\ed
    \end{align}
 In principle, the iterative application of the integration recursion relation \eref{eq:uintrecur} has the potential to provide a complete solution for the generating function. As shown above, we encounter the hypergeometric function in the $n\to n-1$ sector. To obtain an explicit expression for the $n\to n-2$ sector, one needs to deal with the integration of the hypergeometric function, which is quite similar to the pure tensor case. 


However, since we choose to drop the $u_i$ with a certain order, the permutation symmetry does not manifest. Moreover, after solving the function $\mathcal{F}_{n\to\widehat{\boldsymbol{a}}}^{(u_{1})}$, we need to use \eref{chain} to evaluate the sum of complicated productions, which becomes extremely harder when more
propagators being removed. So we may ask if there is a simpler recursion relation in which the permutation symmetry of propagators manifests itself and simplifies the calculation. The answer is yes.
One can find that PDEs in \eref{pdes} is essentially the equation for the electric field $\boldsymbol{E}=-\nabla \phi$ in electromagnetism. As discussed in the former section, after setting
\begin{align}
    F_n=|Q|^{\gamma_n}f_n\co
\end{align}
we have
\begin{align}
    |Q|^{\gamma_{n}}\partial_{u_{i}}f_{n}(\{u\})&=\sum_{\boldsymbol{a}\not=\emptyset}C_{n\to\widehat{\boldsymbol{a}}}^{(H_{i})}F_{n,\widehat{\boldsymbol{a}}}[\{u\}]\co
\end{align}
where we can regard $u_i$ as coordinates in $n$-dimensional space and $f_n$ as the potential. And we define the electric field as 
\begin{align}
 E^i(\{u\})\equiv  |Q|^{-\gamma_n}\sum_{\boldsymbol{a}\not=\emptyset}C_{n\to\widehat{\boldsymbol{a}}}^{(H_{i})}F_{n,\widehat{\boldsymbol{a}}}[\{u\}]\ed
\end{align}
Then we can choose an arbitrary path $\gamma:[0,1]\to \mathbb{R}^n$ from the origin to the final point $(u_1,u_2,\ldots,u_n)$, so that 
\begin{align}
    f_n(\{u\})=|Q_0|^{-\gamma_n}I_n+\int_0^1 \vec{E}(\vec{\gamma}(t))\cdot d\vec{\gamma}(t)\ed
\end{align} 
After choosing the straight line path $x_i(t)=t u_i$,  we can obtain the expression with the wanted permutation symmetry being manifested
\begin{align}
    F_n(\{u\})=\left(\frac{|Q|}{|Q_{0}|}\right)^{\gamma_{n}}I_{n}+|Q|^{\gamma_n}\int_0^1dt~|Q(t)|^{-\gamma_n}\sum_{\boldsymbol{a}\not=\emptyset}\sum_{i}\left(u_iC_{n\to\widehat{\boldsymbol{a}}}^{(H_{i})}(t)\right)F_{n,\widehat{\boldsymbol{a}}}(t)\ed \label{main-recur}
\end{align}
This formula is the main result of this paper. It is an integral recursion relation that enables us to calculate the generating function iteratively. Similar formula is also obtained in \cite{Feng:2024qsa}(equation 6.40). Here we declare that expressing a quantity $A(\{u\})$ as a function of t implies replacing all its instances of $u_{1\leq i \leq n}$  with $u_it$, \emph{i.e.}
\begin{align}
    A(t)\equiv A(\{ut\})\ed
\end{align}
For later use, we also define
\begin{align}
    \overtr{AB}^{(t)}\equiv AQ^{-1}(t)B,~ (\widetilde{AB})^{(t)}\equiv AQ^{*}(t)B\ed
\end{align}
This definition also applies to cases where there are rows and columns of $Q$ removed. We call such a replacement ``scaling". We will meet scaling with other variables, which are not a single $t$ but products in section \ref{general case}. That is just a direct generalization and we will no longer specifically state it.

Taking the contribution for a certain master integral, we have 
\begin{align}
    \mathcal{F}_{n\to\widehat{\boldsymbol{a}}}(\{u\})=\left(\frac{|Q|}{|Q_{0}|}\right)^{\gamma_{n}}\delta_{\boldsymbol{a},\emptyset}+|Q|^{\gamma_{n}}\sum_{\boldsymbol{b}\not=\emptyset,\boldsymbol{b}\subset\boldsymbol{a}}
    \int_{0}^{1}dt|Q(t)|^{-\gamma_n}\left(\sum_{i}u_{i}C_{n\to\widehat{\boldsymbol{b}}}^{(H_{i})}(t)\right)\mathcal{F}_{n,\widehat{\boldsymbol{b}}\to\widehat{\boldsymbol{a}}}(t)\ed
    \label{main_result}
\end{align}
Setting $\boldsymbol{a}=\emptyset$, we get again the same result of $n\to n$ as \eref{eq:ntoncoefficient}. With this equation, we could determine the reduction coefficients recursively: first, we will derive the coefficient of $n\to n-1$ from that of $n\to n$, then $n\to n-2$ with both the results, and so on. Our strategy is to write the integrand as rational expression, and expand it with respect to the integration variable $t$ using Taylor series expansion, and then exchange the order of integration and summation to obtain the final series solution. The advantage of this strategy is that because of the scaling, $t$ will always appear with a $u_{1\leq i\leq n}$, and the reverse is also true except $u_i$ in $\sum_{i}u_{i}C_{n\to\widehat{\boldsymbol{b}}}^{(H_{i})}(t)$ appears alone as it is the direction vector. This fact indicates the coefficient of $t^k$ in the expansion of integrand is the homogeneous polynomial of degree $k+1$ about $\{u\}$. In practical usage, since we are only concerned with the coefficients of $\mathcal{F}_{n\to\widehat{\boldsymbol{a}}}(\{u\})$ about a specific power of $\{u\}$, we can truncate the series solution to a finite number of terms. We will illustrate this point in the later sections.

To do this expansion, we need to understand the dependency of expressions like $|Q|, Q^*_{ij}$ on $\{u_i\}$. \eref{expansion_q} gives one, and the remaining we need to know are just:
\begin{align}
\widetilde{H_iL}=\sum_k \frac{(-1)^{(i+k+1)}}{2}(\widetilde{LL})_{(i;k)}^{(0)}u_k+(\widetilde{H_iL})^{(0)}\co
\end{align}
and
\begin{align}
    ~Q^*_{ib}=\sum_{k,l} (Q^*_{ib})^{(kl)}u_k u_l+\sum_{k} (Q^*_{ib})^{(k)}u_k+(Q^*_0)_{ib}\co
\end{align}
where
\begin{align}
(Q^*_{ib})^{(kl)} &= \frac{(-1)^{i+j+k+l+1+\theta(k-i)+\theta(l-j)}}{4}(1-\delta_{ik})(1-\delta_{jl})(\widetilde{LL})_{(ik;jl)}^{(0)}\nn
(Q^*_{ib})^{(k)} &= \frac{(-1)^{i+j}}{2}\left[(1-\delta_{kj}) (\widetilde{H_{k}  L})^{(0)}_{(i;j)}+ (1-\delta_{ki})(\widetilde{H_{k}L})^{(0)}_{(j;i)}\right]\co
\end{align}
and a useful discovery is $\widetilde{LL}$ does not contain $\{u\}$, which means
\begin{align}
(\widetilde{LL})^{(t)}=\widetilde{LL}=(\widetilde{LL})^{(0)}\ed
\end{align}
These expressions are all polynomials of degrees no larger than 2 about $\{u\}$, which means after scaling, they are all binomials\footnote{In this paper, the term ``binomial" is in general sense, which means it could also be a linear term or a constant.} about $t$. All these relations can be checked using simple matrix determinant operations. 
\subsection{$n\to n-1$}
Now we consider the first non-trivial case. Using \eref{main-recur} and the result of $n\to n$, 
one can find the $n\to n-1$ reduction sector of the higher-pole generating function is 
\begin{align}
\label{n-1 integral}
    {{\cal F}_{n\to\widehat{b}}(\{u\})}=\frac{-(D-n)}{4}|Q|^{\gamma_{n}}\frac{(\widetilde{LL})_{(b)}}{|Q_{0(b)}|^{\gamma_{n{-}1}}}\int_{0}^{1}dt\left(|Q(t)|^{-\gamma_{n}-1}{|Q_{(b)}(t)|}^{\gamma_{n{-}1}-1}\sum_{i=1}^{n} {Q_{ib}^{*}(t)u_i} \right)\ed
\end{align}
Here we define a new vector
\begin{align}
   U=\{u_1,u_2,\ldots,u_n\}\co
\end{align}
thus we can write 
\begin{align}
\sum_{i=1}^{n} {Q_{ib}^{*}u_i}\equiv\widetilde{H_bU}, ~~~\sum_{i=1}^{n} {Q_{ib}^{*}(t)u_i}=(\widetilde{H_bU})^{(t)}\ed
\end{align}
Note that we do the $t$-scaling only in the $Q$ matrix of $(\widetilde{H_bU})^{(t)}$. Using previous expansion relations, we can expand $|Q(t)|,|Q_{(b)}(t)|$ and $(\widetilde{H_bU})^{(t)}$ as binomial about $t$ easily
\begin{align}
\label{expansion1}
    |Q(t)|&\equiv |Q|^{tt}t^2 +|Q|^{t} t +|Q_0|,\nn
    |Q_{(b)}(t)|&\equiv |Q_{(b)}|^{tt}t^2 +|Q_{(b)}|^{t} t +|Q_{0(b)}|,\nn
    (\widetilde{H_bU})^{(t)}&\equiv (Q^*)_b^{t}~t +(Q_0^*)_{b}\ed
\end{align}
Note here we used an identity: $\sum_{i}(Q^*_{ib})^{(kl)}u_ku_lu_i=0$, which can be verified easily. Using the binomial theorem, we could expand the expression $A(t)^\nu\equiv (a t^2+b t+c)^\nu$ as\footnote{Here $a$ could be $0$.}
\begin{align}
\label{bio}
	A(t)^\nu&=\sum_{m=0}^\infty \Big\{\sum_{j=0}^{\lfloor\frac{m}{2}\rfloor}\frac{(\nu-m+j+1)_{(m-j)}}{(m-2j)!~j!}a^jb^{m-2j}c^{\nu-m+j}\Big\} t^m\nn
 &\equiv  \sum_{m=0}^\infty \mathcal{C}_m[A;\nu] t^m\co
\end{align}
where 
\begin{align}
k_{(n)}\equiv  k(k+1)\cdots(k+n-1),~ \forall k, n\in \mathbb{N}\ed
\end{align}
And we also define:
\begin{align}
A_1(t)^{\nu_1}\cdots A_n(t)^{\nu_n}\equiv \sum_{k=0}^{\infty}\mathcal{C}_{k}[A_1(t),\cdots,A_n(t);\nu_1,\cdots,\nu_n] t^k\co
\end{align}
where
\begin{align}
\mathcal{C}_{k}[A_1(t),\cdots,A_n(t);\nu_1,\cdots,\nu_n]\equiv \sum_{k_1+\cdots+k_n=k}\mathcal{C}_{k_1}[A_1(t);\nu_n]\cdots\mathcal{C}_{k_n}[A_n(t);\nu_n]\ed
\end{align}
Actually we can use
the integral
\begin{align}
\label{integral1}
    \int_0^1 dt(|Q|^{-\gamma_n-1}~ t^k)=\frac{1}{|Q_0|^{\gamma_n{+}1}(1+k)}~F_1\left(k+1;\gamma_n+1,\gamma_n+1;k+2;1/x_1,1/x_2\right)\co
\end{align}
where $x_1$ and $x_2$ are the two roots of equation $|Q(t)|=0$ while $F_1$ is Appell's hypergeometric function, to get
\begin{align}
    {{\cal F}_{n\to\widehat{b}}(\{u\})}=&\frac{-(D-n)}{4}|Q|^{\gamma_{n}}\frac{(\widetilde{LL})_{(b)}}{(|Q_{0(b)}|)^{\gamma_{n{-}1}}}\sum_{k=0}^{\infty} \frac{\mathcal{C}_k[|Q(t)|, (\widetilde{H_bU})^{(t)}; -\gamma_n-1, 1]}{|Q_0|^{\gamma_n{+}1}(1+k)}\nn
    &\times F_1\left(k+1;\gamma_n+1,\gamma_n+1;k+2;1/x_1,1/x_2\right)\ed
\end{align}
With some good properties of hypergeometric function and $Q$ matrix, the above summation may be calculated and written into an elegant expression\footnote{The reduction coefficient of $n\to n-1$ can be expressed in a non-series form, involving ordinary hypergeometric function ${}_2F_1(a,b;c;z)$(see Eq(6.35) in \cite{Feng:2024qsa})}. But as previous explained, our main strategy is to expand the whole integrand, so from \eref{n-1 integral} we can get
\begin{align}
\label{1_result}
{{\cal F}_{n\to\widehat{b}}(\{u\})}=\frac{-(D-n)}{4}|Q|^{\gamma_{n}}\frac{(\widetilde{LL})_{(b)}}{|Q_{0(b)}|^{\gamma_{n{-}1}}}\sum_{k=0}^{\infty}\frac{\mathcal{C}_k[|Q(t)|,|Q_{(b)}(t)|, (\widetilde{H_bU})^{(t)}; -\gamma_n-1,\gamma_{n-1}-1, 1]}{k+1}\ed
\end{align}
Note here the $\mathcal{C}_k[\cdot]$ in the summation is a homogeneous polynomial of degree $k+1$ about $\{u\}$, which means if we want to calculate the reduction of $n$-point higher-pole integral with a sum of propagators' powers $k+n$, we can only keep the first $k$ terms in the summation. One can also expand the non-homogeneous term outside the summation in \eref{1_result} to express the whole ${{\cal F}_{n\to\widehat{b}}(\{u\})}$ as a summation of homogeneous polynomials, but for practical usage, this form is fine enough.

To check the consistency of our result, we calculate the coefficient of $u_j$ from \eref{1_result}, and expect it equal to \eref{eq: Ccoefficient}. Choose $k=0$, then
\begin{align}
\label{C0}
\mathcal{C}_0[|Q(t)|,|Q_{(b)}(t)|,& (\widetilde{H_bU})^{(t)}; -\gamma_n-1,\gamma_{n-1}-1, 1]\nn
&=|Q_0|^{-\gamma_n-1}|Q_{0(b)}|^{\gamma_{n-1}-1}\sum_{i=0}^{n}(Q^*_0)_{ib}~u_i\ed
\end{align}
As this expression is already a degree-1 homogeneous polynomial of $\{u\}$, we take $Q$ outside the summation in \eref{1_result} as $Q_0$, then we get the coefficient of $u_j$
\begin{align}
{{\cal F}_{n\to\widehat{b}}(\{u\})}\Big|_{u_j}&=\frac{-(D-n)}{4}\frac{(\widetilde{LL})_{(b)}}{|Q_{0(b)}|}\frac{(Q_0^*)_{jb}}{|Q_{0}|}=\frac{-(D-n)}{4}\overtr{LL}_{(b)}\overtr{H_bH_j}^{(0)}
\end{align}
which is the same as $C^{(H_j)}_{n\to\what{b}}$ substituted its $Q$ with $Q_0$ from \eref{eq: Ccoefficient}.

\subsection{$n\to n-2$}

Using the formula \eref{main_result}, we now try to calculate the coefficient of $n\to n-2$. The summation in the integral now has three terms: one representing the direct reduction path of $n\to n-2$, and the other two representing paths of $n\to n -1\to n-2$. We will explain the paths more detailedly in section \ref{general case}.
\begin{align}
    {\cal F}_{n\to\widehat{b_1b_2}}(\{u\})=&|Q|^{\gamma_{n}}\int_{0}^{1}dt|Q(t)|^{-\gamma_{n}}{\cal F}_{n,\widehat{b_{1}b_{2}}\to\widehat{b_{1}b_{2}}}(t)\sum_i u_{i}C_{n\to\widehat{b_{1}b_{2}}}^{(H_{i})}(t)\nn
    &+\left(|Q|^{\gamma_{n}}\int_{0}^{1}dt|Q(t)|^{-\gamma_{n}}{\cal F}_{n,\widehat{b_{1}}\to\widehat{b_{1}b_{2}}}(t)\sum_i u_{i}C_{n\to\widehat{b_{1}}}^{(H_{i})}(t)+(b_{1}\leftrightarrow b_{2})\right)\ed
\end{align}
The first term can be expanded to
\begin{align}
{\cal F}_{(n\to\widehat{b_1b_2})}^{(1)}=&\frac{D-n+1}{8}|Q|^{\gamma_{n}}\frac{(\widetilde{LL})_{(b_1b_2)}}{|Q_{0(b_1b_2)}|^{\gamma_{n{-}2}}}\nn
&\times\int_{0}^{1}dt|Q(t)|^{-\gamma_{n}{-}1}(\widetilde{H_{b_2}L})_{(b_1)}^{(t)}Q_{(b_1)}(t)|^{{-}1}{|Q_{(b_1b_2)}(t)|}^{\gamma_{n{-}2}{-}1}(\widetilde{H_{b_1}U})^{(t)}+(b_{1}\leftrightarrow b_{2})\ed
\end{align}
Using previous expansion relations, after expanding the integrand, we have
\begin{align}
\label{part1}
{\cal F}_{n\to\widehat{b_1b_2}}^{(1)}&=\frac{D-n+1}{8}|Q|^{\gamma_{n}}\frac{(\widetilde{LL})_{(b_1b_2)}}{|Q_{0(b_1b_2)}|^{\gamma_{n{-}2}}}\sum_{k=0}^{\infty}\Big(\frac{1}{k+1}\nn
&\times\mathcal{C}_k[|Q(t)|,(\widetilde{H_{b_2}L})_{(b_1)}^{(t)},|Q_{(b_1)}(t)|,|Q_{(b_1b_2)}(t)|, (\widetilde{H_{b_1}U})^{(t)}; -\gamma_n-1,1,-1,\gamma_{n-2}-1, 1]\Big)\ed
\end{align}
For the second part, with the result from \eref{1_result}, it can be expanded to
\begin{align}
{\cal F}_{n\to\widehat{b_1b_2}}^{(2)}=&\frac{(n-1-D)(n-D)}{16}|Q|^{\gamma_{n}}\frac{(\widetilde{LL})_{(b_1)}(\widetilde{LL})_{(b_1b_2)}}{(|Q_{0(b_1b_2)}|)^{\gamma_{n{-}2}}}
\int_0^1 dt\Big(|Q(t)|^{-\gamma_{n}{-}1} |Q_{(b_1)}(t)|^{\gamma_{n{-}1}-1}(\widetilde{H_{b_1}U})^{(t)}\nn
&\times\sum_{k=0}^{\infty}\frac{\mathcal{C}_k[|Q_{(b_1)}(t)|,|Q_{(b_1b_2)}(t)|, (\widetilde{H_{b_2}U})_{(b_1)}^{(t)}; -\gamma_{n-1}-1,\gamma_{n-2}-1, 1]~t^{k+1}}{k+1}\Big)\nn
&+(b_{1}\leftrightarrow b_{2})\co
\end{align}
Again, after expanding the integrand, we get
\begin{align}
\label{part2}
{\cal F}_{n\to\widehat{b_1b_2}}^{(2)}=&\frac{(n-1-D)(n-D)}{16}|Q|^{\gamma_{n}}\frac{(\widetilde{LL})_{(b_1)}(\widetilde{LL})_{(b_1b_2)}}{(|Q_{0(b_1b_2)}|)^{\gamma_{n{-}2}}}\sum_{k=1}^{\infty}\sum_{l=0}^{k-1}\Big(\frac{1}{(k-l)(k+1)}\nn
&\times\mathcal{C}_l[|Q_(t)|,|Q_{(b_1)}(t)|, (\widetilde{H_{b_1}U})^{(t)}; -\gamma_{n}-1,\gamma_{n-1}-1, 1]\nn
&\times\mathcal{C}_{k-l-1}[|Q_{(b_1)}(t)|,|Q_{(b_1b_2)}(t)|, (\widetilde{H_{b_2}U})_{(b_1)}^{(t)}; -\gamma_{n-1}-1,\gamma_{n-2}-1, 1]\Big)\nn
&+(b_{1}\leftrightarrow b_{2})\ed
\end{align}
So finally, we can derive
\begin{align}
\label{2parts}
{\cal F}_{n\to\widehat{b_1, b_2}}={\cal F}_{n\to\widehat{b_1, b_2}}^{(1)}+{\cal F}_{n\to\widehat{b_1, b_2}}^{(2)}\ed
\end{align}
And here $\mathcal{C}_k[\cdot]$ in ${\cal F}_{n\to\widehat{b_1, b_2}}^{(1)}$ and $\mathcal{C}_l[\cdot]\mathcal{C}_{k-l-1}[\cdot]$ in ${\cal F}_{n\to\widehat{b_1, b_2}}^{(2)}$ are both homogeneous polynomials of degree $k+1$ about $\{u\}$. Like the case of $n\to n-1$, we can check the rightness by setting $k=0$ and calculating the coefficient of $u_j$. Note in \eref{part2} the summation index $k$ begins from 1, so we just need to calculate ${\cal F}_{n\to\widehat{b_1, b_2}}^{(1)}$. When $k=0$, it is easy to get
\begin{align}
\mathcal{C}_0[|Q(t)|,&(\widetilde{H_{b_2}L})_{(b_1)}^{(t)},|Q_{(b)}(t)|,|Q_{(b_1b_2)}(t)|, (\widetilde{H_{b_1}U})^{(t)}; -\gamma_n-1,1,-1,\gamma_{n-1}-1, 1]\nn
&=|Q_0|^{-\gamma_n-1}(\widetilde{H_{b_2}L})_{(b_1)}^{(0)}|Q_{0(b)}|^{-1}|Q_{0(b_1b_2)}|^{\gamma_{n-2}-1}\sum_i (Q_0^*)_{ib_1}u_i+(b_{1}\leftrightarrow b_{2})\ed
\end{align}
and taking $Q$ to $Q_0$ outside the summation in \eref{part1}, we get
\begin{align}
{\cal F}_{n\to\widehat{b_1, b_2}}\Big|_{u_j}&=\frac{D-n+1}{8}\frac{(\widetilde{LL})_{(b1b2)}(\widetilde{H_{b_2}L})_{(b_1)}^{(0)}(Q_0^*)_{jb_1}}{|Q_{0(b_1b_2)}||Q_{0(b1)}||Q_0|}+(b_{1}\leftrightarrow b_{2})\nn
&=\frac{D-n+1}{8}\overtr{LL}_{(b_1b_2)}\Big(\overtr{H_{b_2}L}_{(b_1)}^{(0)}\overtr{H_{b_1}H_j}^{(0)}+(b_{1}\leftrightarrow b_{2})\Big)\ed
\end{align}
which is consistent with $C^{(H_j)}_{n\to\what{b_1b_2}}$ from \eref{eq: Ccoefficient}.

\subsection{General case}
\label{general case}
Actually, the integration recursion relation \eref{main-recur} enables us to find a general expression for any sector $n\to n-m$ with $m\geq 1$ of the higher-pole generating function. At first, we observe the general pattern of the reduction coefficients
\begin{align}
    C_{n\to\widehat{\boldsymbol{a}}}^{(Z)}=\frac{D-n-1+|\boldsymbol{a}|}{2^{|\boldsymbol{a}|+1}}(-1)^{|\boldsymbol{a}|}(\overline{LL})_{(\boldsymbol{a})}\sum_{\sigma\in S_{|n|}}(\overline{H_{a_{\sigma{1}}}Z})\prod_{i=2}^{|\boldsymbol{a}|}(\overline{H_{a_{\sigma(i)}}L})_{(a_{\sigma(1)},a_{\sigma(2)},\ldots, a_{\sigma(i-1)})}\co \label{CZ}
\end{align}
where $\boldsymbol{a}$ is the list of the propagators we want to remove. And we need to sum over all the element $\sigma$ in $S_n$(the permutation group of $\{1,2,3,\ldots, n\}$), which makes the permutation symmetry of these removed propagators manifest. 
Considering using \eref{main_result} twice, we have
\begin{align}
    \mathcal{F}_{n\to\widehat{\boldsymbol{a}}}(\{u\})&=\sum_{\emptyset\subsetneq\boldsymbol{b}_1\subset\boldsymbol{a}}
    \int_{0}^{1}dt_1{|Q|^{\gamma_{n}}\over |Q|^{\gamma_n}(t_1)}\Big(\sum_{i}u_{i}C_{n\to\widehat{\boldsymbol{b}_1}}^{(H_{i})}(t_1)\Big)\mathcal{F}_{n,\widehat{\boldsymbol{b}_1}\to\widehat{\boldsymbol{a}}}(t_1)\nn
    &=\int_{0}^{1}dt_1{|Q|^{\gamma_{n}}\over |Q|^{\gamma_n}(t_1)}\Big(\sum_{i}u_{i}C_{n\to\widehat{\boldsymbol{a}}}^{(H_{i})}(t_1)\Big)\mathcal{F}_{n,\widehat{\boldsymbol{a}}\to\widehat{\boldsymbol{a}}}(t_1)\nn
    &~~~~+\sum_{\emptyset\subsetneq\boldsymbol{b}_{1}\subsetneq \boldsymbol{b}_{2}\subset\boldsymbol{a}}
    \int_{0}^{1}dt_1dt_2\bigg[{|Q|^{\gamma_{n}}\over |Q|^{\gamma_n}(t_1)}\Big(\sum_{i}u_{i}C_{n\to\widehat{\boldsymbol{b}_1}}^{(H_{i})}(t_1)\Big)\nn
    &~~~~~~~~\times{|Q_{(\boldsymbol{b_1})}|^{\gamma_{n-|\boldsymbol{b}_1|}}\over |Q_{(\boldsymbol{b}_1)}|^{\gamma_{n-|\boldsymbol{b}_1|}}(t_1)}\Big(\sum_{i}u_{i}t_1C_{n,\widehat{\boldsymbol{b}_1}\to\widehat{\boldsymbol{b}_2}}^{(H_{i})}(t_1t_2)\Big)\mathcal{F}_{n,\widehat{\boldsymbol{b}_2}\to\widehat{\boldsymbol{a}}}(t_1t_2)\bigg]\ed
    \label{iterative}
\end{align}
 The first term represents the case $\boldsymbol{b}_1=\boldsymbol{a}$ while the second is the sum of the ones where we have to arrange the task to remove propagators $\boldsymbol{a}$ in at least 2 steps. Actually we can treat the second term using \eref{main_result} iteratively, until we find every reduction path $\emptyset=\boldsymbol{b}_0\subsetneq\boldsymbol{b}_{1}\subsetneq \boldsymbol{b}_{2}\subsetneq \boldsymbol{b}_{3}\cdots\subsetneq \boldsymbol{b}_{j}=\boldsymbol{a}$, where we use $\boldsymbol{b}_r$ to record the propagators have been removed in the $r$-th step. We will denote such a $j$-step path with $\boldsymbol{\mathcal{P}}_j$. So finally we will obtain a multiple integral where the only reduction coefficient appearing is $\mathcal{F}_{n,\widehat{\boldsymbol{a}}\to\widehat{\boldsymbol{a}}}$ which we have known
\begin{align}
    \mathcal{F}_{n,\widehat{\boldsymbol{a}}\to\widehat{\boldsymbol{a}}}(\{u\})=\frac{|Q_{(\boldsymbol{a})}|^{\gamma_{n-|\boldsymbol{a}|}}}{|Q_{0(\boldsymbol{a})}|^{\gamma_{n-|\boldsymbol{a}|}}}\ed
\end{align}
For simplicity, we also define 
\begin{align}
      T_r\equiv \prod_{s=1}^{r}t_{s},~T_{0}\equiv 1\co
\end{align}
then the mutiple integral can be written as
\begin{align}
    {\cal F}_{n\to\widehat{\boldsymbol{a}}}(\{u\})=&\sum_{j=1}^{|\boldsymbol{a}|}\sum_{\boldsymbol{\mathcal{P}}_j}\int_{0}^{1}\left[\prod_{i=1}^{j}dt_{i}\right]\frac{|Q_{(\boldsymbol{a})}\left(T_j\right)|^{\gamma_{n-|\boldsymbol{a}|}}}{|Q_{0(\boldsymbol{a})}|^{\gamma_{n-|\boldsymbol{a}|}}}\nn
    &\times \frac{1}{T_j}\prod_{r=1}^{j}\left[\frac{|Q_{(\boldsymbol{b}_{r-1})}\left(T_{r-1}\right)|^{\gamma_{n-|\boldsymbol{b}_{r-1}|}}}{|Q_{(\boldsymbol{b}_{r-1})}\left(T_{r}\right)|^{\gamma_{n-|\boldsymbol{b}_{r-1}|}}}C_{\widehat{\boldsymbol{b}_{r-1}}\to\widehat{\boldsymbol{b}_{r}}}^{(U)}\left(T_r\right)\right]\ed \label{n-k}
\end{align}
Here we have used
\begin{align}
    C_{\widehat{\boldsymbol{b}_{r-1}}\to\widehat{\boldsymbol{b}_{r}}}^{(U)}\left(T_r\right)\equiv\sum_{i}u_iT_r C_{n,\widehat{\boldsymbol{b}_{r-1}}\to\widehat{\boldsymbol{b}_{r}}}^{(H_i)}\left(T_r\right)\ed
    \label{C(U)}
\end{align}
Note that in  \eref{C(U)}, we have introduced an additional factor $t_r$ in the scaling of $u_i$ at each step. Since $U=[u_1,\cdots,u_n]$ serves as the direction vector for each $t_r$-integral and should not be scaled by $t_r$, we must divide it. This is the reason why the term $1/T_j$ arises.

Substituting \eref{CZ} into \eref{n-k} and taking quantities without $T_{1\leq r\leq j}$ out of the integral, we have
\begin{align}\label{eq:n2k}
    {\cal F}_{n\to\widehat{\boldsymbol{a}}}(\{u\})=&\sum_{j=1}^{|\boldsymbol{a}|}\sum_{\boldsymbol{\mathcal{P}}_j}\frac{(-1)^{|\boldsymbol{a}|}}{2^{|\boldsymbol{a}|+j}}\frac{|Q|^{\gamma_{n}}}{|Q_{0(\boldsymbol{a})}|^{\gamma_{n-|\boldsymbol{a}|}}}\int_{0}^{1}\left[\prod_{i=1}^{j}dt_{i}\right]\frac{|Q_{(\boldsymbol{a})}\left(T_{j}\right)|^{\gamma_{n-|\boldsymbol{a}|}}}{T_j|Q(T_1)|^{\gamma_{n}}}\nn
    &\quad\times\prod_{r=2}^{j}\bigg[\frac{|Q_{(\boldsymbol{b}_{r-1})}\left(T_{r-1}\right)|^{\gamma_{n-|\boldsymbol{b}_{r-1}|}}}{|Q_{(\boldsymbol{b}_{r-1})}\left(T_{r}\right)|^{\gamma_{n-|\boldsymbol{b}_{r-1}|}}}\bigg]\prod_{r=1}^{j}\bigg[(D-n-1+|\boldsymbol{b}_{r}|)\overtr{LL}_{(\boldsymbol{b}_r)}^{(T_r)}\nn
    &\quad\quad\times \sum_{\sigma\in S_{|\boldsymbol{c_r}|}}\Big[\overtr{H_{[\boldsymbol{c}_r]_{\sigma(1)}}U}_{(\boldsymbol{b}_{r-1})}(T_r)\prod_{i=2}^{|\boldsymbol{c}_r|}\overtr{H_{[\sigma\boldsymbol{c}_r]_i}L}_{(\boldsymbol{b}_{r-1}\cup[\boldsymbol{c}_r]_{\{\sigma(1),\ldots,\sigma(i-1)\}})}^{(T_r)}\Big]\bigg]\co
\end{align}
Here $\boldsymbol{c}_{r}=\boldsymbol{b}_{r}-\boldsymbol{b}_{r-1}$ and $[\boldsymbol{c}_r]_{\sigma(i)}$ denotes the $i$-th element of the list $\boldsymbol{c}_r$ after applying the permutation operator $\sigma$, while $[\boldsymbol{c}_r]_{\{\sigma(1),\ldots,\sigma(i-1)\}}$ represents the first $i-1$ elements of the list following the permutation operation. Note again that the scaling in $\overtr{H_{[\boldsymbol{c}_r]_{\sigma(1)}}U}_{(\boldsymbol{b}_{r-1})}(T_r)$ does not only occur in the omitted $Q$ matrix but also in $U$, which makes it have an additional factor $T_r$ than $\overtr{H_{[\boldsymbol{c}_r]_{\sigma(1)}}U}_{(\boldsymbol{b}_{r-1})}^{(T_r)}$.
After reorganizing it, we get
\begin{align}\label{eq:n2k_final}
    {\cal F}_{n\to\widehat{\boldsymbol{a}}}(\{u\})=&\sum_{j=1}^{|\boldsymbol{a}|}\sum_{\boldsymbol{\mathcal{P}}_j}\frac{(-1)^{|\boldsymbol{a}|}}{2^{|\boldsymbol{a}|+j}}\frac{|Q|^{\gamma_{n}}}{|Q_{0(\boldsymbol{a})}|^{\gamma_{n-|\boldsymbol{a}|}}}\int_{0}^{1}\left[\prod_{i=1}^{j}dt_{i}\right]\frac{1}{T_j}\prod_{r=1}^{j}A^{\boldsymbol{b}_{r-1}, \boldsymbol{b}_{r}}(T_r)\co
\end{align}
where
\begin{align}
A^{\boldsymbol{b}_{r-1}, \boldsymbol{b}_{r}}(T_r)&=\frac{|Q_{(\boldsymbol{b}_{r})}\left(T_{r}\right)|^{\gamma_{n-|\boldsymbol{b}_{r}|}}}{|Q_{(\boldsymbol{b}_{r-1})}\left(T_{r}\right)|^{\gamma_{n-|\boldsymbol{b}_{r-1}|}}}(D-n-1+|\boldsymbol{b}_{r}|)\overtr{LL}_{(\boldsymbol{b}_r)}^{(T_r)}\nn
    &\quad\quad\times \sum_{\sigma\in S_{|\boldsymbol{c_r}|}}\Big[T_r\overtr{H_{[\boldsymbol{c}_r]_{\sigma(1)}}U}_{(\boldsymbol{b}_{r-1})}^{(T_r)}\prod_{i=2}^{|\boldsymbol{c}_r|}\overtr{H_{[\sigma\boldsymbol{c}_r]_i}L}_{(\boldsymbol{b}_{r-1}\cup[\boldsymbol{c}_r]_{\{\sigma(1),\ldots,\sigma(i-1)\}})}^{(T_r)}\Big]\ed
\end{align}
Note that all quantities appearing in $\mathcal{A}^{\boldsymbol{b}_{r-1}, \boldsymbol{b}_{r}}(T_r)$ can be written into a product of binomials of $T_{r}$ with some powers.\footnote{For example, $\overtr{LL}_{(\boldsymbol{b}_{r})}^{(T_r)}=|Q_{(\boldsymbol{b}_{r})}(T_r)|^{-1}(\widetilde{LL})_{(\boldsymbol{b}_{r})}^{(T_r)}$. $|Q_{(\boldsymbol{b}_{r})}(T_r)|$ is a binomial of $T_r$ while $(\widetilde{LL})_{(\boldsymbol{b}_{r})}^{(T_r)} = (\widetilde{LL})_{(\boldsymbol{b}_{r})}^{(0)}$ is a constant about $T_r$.}. So we can, as in the previous cases of $n\to n-1$ and $n\to n-2$, use the expansion relations about $Q$ matrix previously listed to expand it into a power series of $T_{r}$ 
\begin{align}
A^{\boldsymbol{b}_{r-1}, \boldsymbol{b}_{r}}(T_r)&=\sum_{k=1}^{\infty} \mathcal{A}_k^{\boldsymbol{b}_{r-1}, \boldsymbol{b}_{r}}(\{u\})T_r^k\ed
\end{align}
The summation begins from $k=1$ according to the scaling of $U$.
So finally, the form of the integral in \eref{eq:n2k_final} will be
\begin{align}
\label{intgral_expansion}
\int_{0}^{1}\left[\prod_{i=1}^{j}dt_{i}\right]{1\over T_j}\sum_{\{k\}=1}\prod_{r=1}^{j} \mathcal{A}_{k_r}^{\boldsymbol{b}_{r-1}, \boldsymbol{b}_{r}}(\{u\})T_r^{k_r}\ed
\end{align}
And the final result is 
\begin{align}
\label{general result}
{\cal F}_{n\to\widehat{\boldsymbol{a}}}=&\sum_{j=1}^{|\boldsymbol{a}|}\sum_{\boldsymbol{\mathcal{P}}_j}\frac{(-1)^{|\boldsymbol{a}|}}{2^{|\boldsymbol{a}|+j}}\frac{|Q|^{\gamma_{n}}}{|Q_{0(\boldsymbol{a})}|^{\gamma_{n-|\boldsymbol{a}|}}}\sum_{\{k\}=1}\prod_{r=1}^{j}\frac{\mathcal{A}^{\boldsymbol{b}_{r-1}, \boldsymbol{b}_{r}}_{k_r}(\{u\})}{\sum_{m=r}^{j}k_m}
\end{align}
Note that in $A^{\boldsymbol{b}_{r-1}, \boldsymbol{b}_{r}}(T_r)$, every $u_{1\leq i\leq j}$ appears with $T_{r}$ in a product form, and the reverse is also true. So $\mathcal{A}^{\boldsymbol{b}_{r-1}, \boldsymbol{b}_{r}}_{k_r}(\{u\})$ is a homogeneous polynomial of degree $k_r$ about $\{u\}$. Thus to derive the reduction of $n$-point integrals with a sum of the propagators' powers $n+v$, we only need to keep the terms with $\sum_{r=1}^{j}k_r\leq v$ in the summation of every possible reduction path. 
So, when $v=1$, only the path with length $j=1$ ($\boldsymbol{b}_1=\boldsymbol{a}$) in \eref{general result} is necessary and we can only calculate $\mathcal{A}_1^{\emptyset, \boldsymbol{a}}$ to determine the coefficient of $u_{1\leq i\leq n}$. This is equivalent to, in \eref{n-k}, remove other irrelevant paths in the summation and set $T_1=0$ in the integrand of the only left path. So we get
\begin{align}
{\cal F}_{n\to\widehat{\boldsymbol{a}}}(\{u\})=\left(\frac{|Q_{(\boldsymbol{a})}|}{|Q_{0(\boldsymbol{a})}|}\right)^{\gamma_{n-|\boldsymbol{a}|}}\sum_i u_i C_{n\to\widehat{\boldsymbol{a}}}^{(H_i)}(0)\ed
\end{align}
Taking the coefficient of $u_l$ and we get
\begin{align}
{\cal F}_{n\to\widehat{\boldsymbol{a}}}(\{u\})\vert_{u_l}=C_{n\to\widehat{\boldsymbol{a}}}^{(H_l)}(0)\co
\end{align}
which is exactly the same as \eref{CZ} where $Q$ is set to $Q_0$ and $Z=H_l$ . And this is exactly the consistency we have shown in the cases of $n\to n-1$ and $n \to n-2$.\footnote{For the case of $n\to n$, the consistency can be checked very easily.}

In this section, we outline our strategy without providing the detailed expression of $\mathcal{A}^{\boldsymbol{}}_{k_1,\cdots,k_j}({u})$ due to its complexity. In the future, we will explore the possibility of uncovering any hidden symmetries in the $Q$ matrix to facilitate simplification.

\section{Examples}\label{sec:example}

In this section, we will give some examples to illustrate how to use our methods. We denote the reduction coefficient of $I_{\{v_1,\ldots,v_n\}} \to I_{n,\widehat{\boldsymbol{a}}}$ with $C_{\{v_1,\ldots,v_n\}\to\widehat{\boldsymbol{a}}}$\footnote{Unlike \eref{eq:one-loop}, here the $i$-th propagator's power is $v_i$ not $v_i+1$.}. For the coefficient $C_{\{v_1,\ldots,v_n\}\to\widehat{\boldsymbol{\emptyset}}}$, as \eqref{eq:ntoncoefficient} is a totally analytic result, what we have to do is just expand it as series of $\{u\}$, so we just illustrate it in the case of tadpole.
 
\subsection{Tadpoles}

 \begin{align}
 \gamma_1=\frac{D-2}{2},~
     Q=\left[
     \begin{array}{c}
     m_1^2+u_1 
     \end{array}
     \right],~
      Q^{*}=\left[
     \begin{array}{c}
      1
     \end{array}
     \right]\co   
\end{align}
It is easy to find using \eqref{eq:ntoncoefficient}
\begin{align}
    {\cal F}_1[{\{u\}}] &= \left(\frac{m_1^2+u_1}{m_1^2}\right)^{\gamma_{1}}=\sum_{n=0}^{\infty} {\gamma_1 \choose n} \frac{u_1^n}{m_1^{2n}}\co
\end{align}
where 
\begin{align}
    {\gamma_1 \choose n}\equiv \frac{\gamma_1(\gamma_1-1)\ldots(\gamma_1-n+1)}{n!}\ed
\end{align}
It means $C_{\{n+1\}\to\widehat{\emptyset}}={\gamma_1 \choose n} \frac{1}{m_1^{2n}}$. The validity of the formula can be readily verified by employing IBP identity.

\subsection{Bubbles}
Now we turn to deal with the first nontrivial case, \emph{i.e.}, the bubbles, where
\begin{align}
\gamma_2&=\frac{D-3}{2},~ Q=\left(
\begin{array}{cc}
 m_{1}^2+u_{1} & \frac{1}{2} \left(m_{1}^2+m_{2}^2-s_{11}+u_{1}+u_{2}\right) \\
\frac{1}{2} \left(m_{1}^2+m_{2}^2-s_{11}+u_{1}+u_{2}\right) & m_{2}^2+u_{2} \\
\end{array}
\right)\co\nn
    |Q_0|&=-\frac{1}{4} \left(m_{1}^2-2 m_{1} m_{2}+m_{2}^2-s_{11}\right) \left(m_{1}^2+2 m_{1} m_{2}+m_{2}^2-s_{11}\right)\ed
\end{align}
Then
\begin{align}
    Q(t)&=\left(
\begin{array}{cc}
 m_{1}^2+u_{1}t & \frac{1}{2} \left(m_{1}^2+m_{2}^2-s_{11}+u_{1}t+u_{2}t\right) \\
\frac{1}{2} \left(m_{1}^2+m_{2}^2-s_{11}+u_{1}t+u_{2}t\right) & m_{2}^2+u_{2} \\
\end{array}
\right)\co\nn
(\widetilde{H_2U})^{(t)}&=H_2\cdot Q^{*}(t)\cdot U= \begin{pmatrix}
    0 & 1
\end{pmatrix}Q^{*}\left(
\begin{array}{c}
 u_1 \\
 u_2 \\
\end{array}
\right)\ed
\end{align}
The $2\to 2$ sector is simply given by the formula \eref{eq:ntoncoefficient}, so here we just discuss the $2\to 1$ sector, \emph{i.e.}, the reduction to tadpole, to check the formula for $n\to n-1$ sector. Here we calculate the reduction coefficients for cases of $\{v_1v_2\}\to\widehat{2}$ with $v_1+v_2\le 3$, so we need to calculate  $\mathcal{C}_0[\cdot], \mathcal{C}_1[\cdot]$ and  $\mathcal{C}_2[\cdot]$ in \eref{1_result}.
Using \eref{C0}, we have
\begin{align}
&\mathcal{C}_0[|Q(t)|,|Q_{(2)}(t)|, (\widetilde{H_2U})^{(t)}; -\gamma_2-1,\gamma_{1}-1, 1]
=|Q_0|^{-\gamma_2-1}|Q_{0(2)}|^{\gamma_{2-1}-1}\sum_{i=0}^{2}(Q^*_0)_{i2}u_i\nn
&=-\frac{1}{2} m_1^{D-4} \left(|Q_0|\right){}^{\frac{1-D}{2}}
   \left(\left(m_1^2 +m_2^2-s_{11}\right){\color{orange}u_1}-2m_1^2 {\color{orange}u_2}\right)\co
\end{align}
\allowdisplaybreaks
and for the others, we have
\begin{align}
&\mathcal{C}_1[|Q(t)|,|Q_{(2)}(t)|, (\widetilde{H_2U})^{(t)}; -\gamma_2-1,\gamma_{1}-1, 1]=\frac{1}{16}m_1^{D-6}|Q_0|{}^{-\frac{D+1}{2}}\times\nn
& \bigg(\Big[m_1^4 \left((D+4) s_{11}-D m_2^2\right)+m_1^2 \left(m_2^2-s_{11}\right) \left((D+4) m_2^2-(D-8) s_{11}\right)\nn
&~~~~~~~~+(D-4) \left(m_2^2-s_{11}\right){}^3 m_1^6\Big]{\color{orange}u_1^2} -\Big[4 (D-1)m_1^4(m_1^2-m_2^2+s_{11})\Big]{\color{orange}u_2^2}\nn
&~~~~+4m_1^2\Big[(2-D) \left(m_2^2-s_{11}\right){}^2+D m_1^4-2 m_1^2 \left(m_2^2+s_{11}\right)\Big]{\color{orange}u_1u_2}\Bigg)\co\nn
&\mathcal{C}_2[|Q(t)|,|Q_{(2)}(t)|, (\widetilde{H_2U})^{(t)}; -\gamma_2-1,\gamma_{1}-1, 1]=-\frac{1}{256}m_1^{D-8}|Q_0|^{-\frac{D+3}{2}}\times\nn
&\bigg(\Big[2 D \big(4 m_1^6 \left(m_2^2 s_{11}-2 m_2^4+s_{11}^2\right)-5 \left(m_2^2-s_{11}\right){}^5+4 m_1^4 \left(m_2^2-s_{11}\right){}^2 \left(m_2^2+2 s_{11}\right)+m_1^{10}\nn
&~~~~~~~~+m_1^2 \left(m_2^2-s_{11}\right){}^3 \left(7 m_2^2+13 s_{11}\right)+m_1^8 \left(m_2^2-5 s_{11}\right)\big)-8 \big(3 m_1^6 \left(m_2^4-3 s_{11}^2\right)+2 m_1^8 s_{11}\nn
&~~~~~~~~+m_1^4 \left(3 m_2^4 s_{11}-9 m_2^2 s_{11}^2-9 m_2^6+15 s_{11}^3\right)+m_1^2 \left(m_2^2-s_{11}\right){}^3\left(9 m_2^2+11 s_{11}\right) \nn
&~~~~~~~~-3 \left(m_2^2-s_{11}\right){}^5\big)+D^2 \left(m_1^2-m_2^2+s_{11}\right){}^2 \left(m_1^2+m_2^2-s_{11}\right){}^3\Big]{\color{orange}u_1^3}\nn
&~~~~+6m_1^2\Big[8 \big(m_1^6 \left(m_2^2+s_{11}\right)-3 m_1^4 \left(m_2^4+s_{11}^2\right)+3 m_1^2 \left(m_2^2-s_{11}\right){}^2 \left(m_2^2+s_{11}\right)-\left(m_2^2-s_{11}\right)^4\big)\nn
&~~~~~~~~-D^2 \left(m_1^4-\left(m_2^2-s_{11}\right){}^2\right){}^2-2 D (2 m_1^6 \left(m_2^2+s_{11}\right)+2 m_1^4 \left(m_2^2-s_{11}\right){}^2+3 \left(m_2^2-s_{11}\right)^4\nn
&~~~~~~~~-6 m_1^2 \left(m_2^2-s_{11}\right){}^2\left(m_2^2+s_{11}\right)-m_1^8)\Big]{\color{orange}u_1^2u_2}\nn
&~~~~+12(D-1)m_1^4\Big[m_1^4\left((D+2) s_{11}-(D+6) m_2^2\right)-m_1^2 \left(m_2^2-s_{11}\right) \left((D-6) m_2^2-(D+6) s_{11}\right)\nn
&~~~~~~~~+(D-2) \left(m_2^2-s_{11}\right){}^3+(D+2) m_1^6\Big]{\color{orange}u_1u_2^2}\nn
&~~~~-8(D-1)m_1^6\Big[D \left(m_1^2-m_2^2+s_{11}\right){}^2+4 m_1^2 s_{11}\Big]{\color{orange}u_2^3}\bigg)\ed
\end{align}
The form of $\mathcal{C}_k[\cdot]$ may be complicated, but we could see they are homogeneous polynomials  of degree $k + 1$ about $\{u\}$ clearly. And finally we can get
\begin{align}
&{\cal F}_{2\to\widehat{2}}(\{u\})\nn
&=\frac{-(D-2)}{4}|Q|^{\gamma_{2}}\frac{(\widetilde{LL})_{(2)}}{|Q_{0(2)}|^{\gamma_1}}\left(\sum_{k=0}^{2}\frac{\mathcal{C}_k[|Q(t)|,|Q_{(2)}(t)|, (\widetilde{H_2U})^{(t)}; -\gamma_2-1,\gamma_{1}-1, 1]}{k+1}+\mathcal{O}(\{u\}^4)\right)\nn
&=\frac{(D-2)(m_1^2+m_2^2-s_{11})}{8m_1^2|Q_0|}{\color{orange}u_1}-\frac{D-2}{4|Q_0|}{\color{orange}u_2}-\frac{(D-5) (D-2) \left(m_1^2-m_2^2+s_{11}\right)}{32 |Q_0|^2}{\color{orange}u_2^2}\nn
&~~~~+\frac{(D-2)}{32 m_{1}^2 |Q_0|^2} \left[(2 D-9) m_{1}^4-2 (D-4) m_{1}^2 \left(m_{2}^2+s_{11}\right)+\left(m_{2}^2-s_{11}\right)^2\right]{\color{orange}u_1u_2}\nn
&~~~~+\frac{D-2}{{128 |Q_0|^2 m_1^4}} \Big[m_1^4 \left(D m_2^2+7 (D-4) s_{11}\right)+m_1^2 \left(m_2^2-s_{11}\right) \left((3 D-16) m_2^2+5 (D-4) s_{11}\right)\nn
&~~~~~~~~-(D-4)
\left(m_2^2-s_{11}\right){}^3-3 (D-4) m_1^6\Big]{\color{orange}u_1^2}\nn
&~~~~-\frac{D-2}{3072|Q_0|^3 m_1^6} \Big[2 m_1^4 \left(m_2^2-s_{11}\right)\big(\left(8 D^2-98 D+276\right) m_2^2 s_{11}+\left(11 D^2-104 D+240\right) s_{11}^2\nn
&~~~~~~~~+(5 D^2-62 D+180) m_2^4\big)+m_1^8 \big(\left(11 D^2-86 D+144\right) m_2^2+5 \left(5 D^2-46 D+104\right) s_{11}\big)\nn
&~~~~~~~~-(7 D^2-64 D+144) m_1^{10}-2 m_1^6 \big((8 D^2-86D+216) m_2^2 s_{11}-\left(D^2-28 D+120\right) m_2^4\nn
&~~~~~~~~+\left(17 D^2-158 D+360\right) s_{11}^2\big)+(D-4) m_1^2 \left(m_2^2-s_{11}\right){}^3\left((5 D-36)m_2^2-(7 D-40) s_{11}\right)\nn
&~~~~~~~~-\left(D^2-10D+24\right) \left(m_2^2-s_{11}\right){}^5\Big]{\color{orange}u_1^3}\nn
&~~~~-\frac{(D-2)}{512|Q_0|^3 m_1^4} \Big[(D-4) \left(m_2^2-s_{11}\right){}^4-4 m_1^2
\left(m_2^2-s_{11}\right){}^2 \left((D-6) m_2^2+(2 D-9) s_{11}\right)\nn
&~~~~~~~~+2 m_1^4 \big((4 D^2-50D+144) m_2^2 s_{11}+\left(2 D^2-17 D+30\right) m_2^4+\left(2 D^2-13 D+18\right) s_{11}^2\big)\nn
&~~~~~~~~-4 m_1^6 \left(\left(2 D^2-19 D+44\right) m_2^2+\left(2 D^2-18 D+41\right) s_{11}\right)+\left(4 D^2-39 D+96\right) m_1^8\Big]{\color{orange}u_1^2u_2}\nn
&~~~~+\frac{D-2}{128 |Q_0|^3
m_1^2} \Big[\left(D^2-10 D+26\right) m_1^6-m_1^4 \left(\left(2 D^2-20 D+51\right) m_2^2-(17-4 D) s_{11}\right)\nn
&~~~~~~~~+m_1^2 \left(m_2^2-s_{11}\right) \big((D^2-10
D+24) m_2^2+\left(D^2-14 D+42\right) s_{11}\big)+\left(m_2^2-s_{11}\right){}^3\Big]{\color{orange}u_1u_2^2}\nn
&~~~~-\frac{D-2}{384|Q_0|^3} \Big[\left(D^2-10 D+27\right)\left(m_2^2-s_{11}\right){}^2+\left(D^2-10 D+27\right) m_1^4\nn
&~~~~~~~~+m_1^2 \big(2 \left(D^2-14 D+43\right) s_{11}-2 \left(D^2-10 D+27\right) m_2^2\big)\Big]{\color{orange}u_2^3}+\mathcal{O}[\{u\}^4]\ed
\end{align}
With this expansion, we could get the reduction coefficient $C_{v_1,v_2\to\widehat{2}}$ easily by
\begin{align}
C_{\{v_1,v_2\}\to\widehat{2}}={\cal F}_{2\to\widehat{2}}(\{u\})\Big|_{u_1^{v_1-1}u_2^{v_2-1}}\ed
\end{align}
   
\subsection{Triangles}
The calculation of bubbles give a check for the formulas of $n\to n-1$ sector. To check the derivations of $n\to n-2$ sector, we take the reduction of triangles as a benchmark. 
When $n=3$, the analytic result will be quite complicated to display. So here we will use the numerical method. We will take $D=6$ and set
\bea
\left(\begin{array}{cccccc}
    m^2_1,&m^2_2,&m^2_3,&s_{11},&s_{12},&s_{22}
\end{array}\right)&=&
\left(\begin{array}{cccccc}
    2, &\frac{5}{3}, & \frac{11}{7}, &\frac{13}{17}, &\frac{19}{23}, &\frac{29}{31}
\end{array}\right)\co
\eea
so we have
\begin{align}
Q=
\left(
\begin{array}{ccc}
 u_1+2 & \frac{1}{2} \left(u_1+u_2+\frac{148}{51}\right) & \frac{1}{2} \left(u_1+u_3+\frac{572}{217}\right) \\
 \frac{1}{2} \left(u_1+u_2+\frac{148}{51}\right) &  u_2+\frac{5}{3} & \frac{1}{2} \left(u_2+u_3+\frac{812006}{254541}\right) \\
 \frac{1}{2} \left(u_1+u_3+\frac{572}{217}\right) & \frac{1}{2} \left(u_2+u_3+\frac{812006}{254541}\right) & u_3+\frac{11}{7} \\
\end{array}
\right)\ed
\end{align}
We are going to study the reduction of $\{v_1v_2v_3\}\to\widehat{23}$ and $\{v_1v_2v_3\}\to\widehat{3}$ where $v_1+v_2+v_3\le 3$, so we need to calculate $\mathcal{C}_{k\le 2}[\cdot]$ in \eref{1_result}, $\mathcal{C}_{k\le 2}[\cdot]$ in \eref{part1} and $\mathcal{C}_{k\le 1}[\cdot]$ in \eref{part2}. We list partial results, while the remaining ones are can be obtained by using permutations of propagators or direct calculation.
\begin{align}
\mathcal{C}_{0}[|Q_{(3)}(t)|,&|Q_{(23)}(t)|, (\widetilde{H_{2}U})_{(3)}^{(t)}; -\gamma_{2}-1,\gamma_{1}-1, 1]=-\frac{6765201 \left(37 {\color{orange}u_1}-51 {\color{orange}u_2}\right)}{2550409 \sqrt{3194}}\co\nn\nn
\mathcal{C}_0[|Q(t)|,&(\widetilde{H_{2}L})_{(3)}^{(t)},|Q_{(3)}(t)|,|Q_{(23)}(t)|, (\widetilde{H_{3}U})^{(t)}; -\gamma_3-1,1,-1,\gamma_{1}-1, 1]\nn
&=\frac{115443976544836947 }{1840431340803447943994}\left(764296{\color{orange} u_1}-8293467 {\color{orange} u_2}+7970627 {\color{orange} u_3}\right)\co\nn\nn
\mathcal{C}_0[|Q_(t)|,&|Q_{(3)}(t)|, (\widetilde{H_{3}U})^{(t)}; -\gamma_{3}-1,\gamma_{2}-1, 1]\nn
&=\frac{6340636927821 \sqrt{\frac{1597}{2}}}{2304860789985532804}\left(764296 {\color{orange}u_1}-8293467 {\color{orange}u_2}+7970627 {\color{orange}u_3}\right)\co\nn\nn
\mathcal{C}_{1}[|Q_{(3)}(t)|,&|Q_{(23)}(t)|, (\widetilde{H_{2}U})_{(3)}^{(t)}; -\gamma_{2}-1,\gamma_{1}-1, 1]\nn
&=-\frac{6765201 }{8146006346 \sqrt{3194}}\left(47920 {\color{orange}u_1^2}-182733 {\color{orange}u_2 u_1}+182070 {\color{orange}u_2^2}\right)\co\nn\nn
\mathcal{C}_1[|Q_(t)|,&|Q_{(3)}(t)|, (\widetilde{H_{3}U})^{(t)}; -\gamma_{3}-1,\gamma_{2}-1, 1]\nn
&=-\frac{323372483318871}{5588200319839650693944988502424 \sqrt{3194}}\times\nn
&~~~~\big(49086274545506577232 {\color{orange}u_1^2}-474745150216609836875 {\color{orange}u_2 u_1}\nn
&~~~~+443137309921225775667 {\color{orange}u_3 u_1}+1647407696285231394036 {\color{orange} u_2^2}\nn
&~~~~+1756455419948917569056 {\color{orange}u_3^2}-3384646622045561345284 {\color{orange}u_2 u_3}\big)\co\nn\nn
\mathcal{C}_1[|Q(t)|,&(\widetilde{H_{2}L})_{(3)}^{(t)},|Q_{(3)}(t)|,|Q_{(23)}(t)|, (\widetilde{H_{3}U})^{(t)}; -\gamma_3-1,1,-1,\gamma_{1}-1, 1]\nn
&=\frac{16491996649262421}{17848711821567844316460293276742256}\times\nn &~~~~\big(15634043107119145336 {\color{orange} u_1^2}+2170804542188032965837 {\color{orange} u_2 u_3}\nn
&~~~~-582748478980851218805 {\color{orange} u_2^2}-1570581302246120406144 {\color{orange}u_3^2}\nn
&~~~~-303465123978001661625 {\color{orange}u_2 u_1}+224535161787512965865 {\color{orange}u_3 u_1}\big)\co\nn\nn
\mathcal{C}_2[|Q_(t)|,&|Q_{(3)}(t)|, (\widetilde{H_{3}U})^{(t)}; -\gamma_{3}-1,\gamma_{2}-1, 1]\nn
&=\frac{970117449956613}{216780001293540937793009493928203888683010304 \sqrt{3194}}\times\nn
&~~~~\big(610343452223399406562328833257200 {\color{orange}u_1^3}\nn
&~~~~-8883882097861222541752787969357154 {\color{orange}u_2 u_1^2}\nn
&~~~~+8214071122826216852908992869579922 {\color{orange}u_3 u_1^2}\nn
&~~~~-116804839905245356700256666099636012 {\color{orange}u_2^3}\nn
&~~~~+38572391477752033080662959586725440 {\color{orange}u_3^2 u_1}\nn
&~~~~+113285484922745387891133994971258472 {\color{orange}u_3^3}\nn
&~~~~+44869641073761153147308721302689464 {\color{orange}u_2^2 u_1}\nn
&~~~~-338390228015827841462089719854252280 {\color{orange}u_2 u_3^2}\nn
&~~~~-82974427141032740304315639428660568 {\color{orange}u_2 u_3 u_1}\nn
&~~~~+341933431509497314813214944611274908 {\color{orange}u_2^2 u_3}\big)\co\nn\nn
\mathcal{C}_2[|Q(t)|,&(\widetilde{H_{2}L})_{(3)}^{(t)},|Q_{(3)}(t)|,|Q_{(23)}(t)|, (\widetilde{H_{3}U})^{(t)}; -\gamma_3-1,1,-1,\gamma_{1}-1, 1]\nn
&=\frac{841091829112383471}{86549415516446219413859040450835402556691863872}\times\nn
&~~~~\big(1429890409156841053503217612440 {\color{orange} u_1^3}\nn
&~~~~-36719305889546962769091591188935 {\color{orange} u_2 u_1^2}\nn
&~~~~+33776166259072545424478917685431 {\color{orange} u_3 u_1^2}\nn
&~~~~+155976321730388513601110095999743 {\color{orange} u_2^2 u_1}\nn
&~~~~+82287584050096116407126068194656 {\color{orange} u_3^2 u_1}\nn
&~~~~-231865398695597435632244677422791 {\color{orange} u_2 u_3 u_1}\nn
&~~~~-615594470932963710628146835122420 {\color{orange} u_2^3}\nn
&~~~~+744832555847731103855295521100948 {\color{orange} u_3^3}\nn
&~~~~-1988714732114963547975732169793628 {\color{orange} u_2 u_3^2}\nn
&~~~~+1859526920452792002296104482967068 {\color{orange} u_2^2 u_3}\big)\ed
\end{align}
And finally, using\eref{1_result}, \eref{part1} and \eref{part2}, keeping the first few terms with an order of $\{u\}$ less than or equal to 3 in the summations, we get
\begin{align}
\mathcal{F}_{3\to \widehat{3}}(\{u\})&=\frac{29781297}{1201626279799598302052811147776}\times\nn
&~~~~\big(5594402172099529552296 {\color{orange}u_1^3}-82263493881535179932199 {\color{orange}u_2 u_1^2}\nn
&~~~~+70690227735613909541391 {\color{orange}u_3 u_1^2}+5121436198492679794176 {\color{orange}u_1^2}\nn
&~~~~+309638117961029004973020 {\color{orange}u_2^2 u_1}+193157394013560093887952 {\color{orange}u_3^2 u_1}\nn
&~~~~+18379528497296359856520 {\color{orange}u_2 u_1}-492796615141101825475740 {\color{orange}u_2 u_3 u_1}\nn
&~~~~-43365324908689513165128 {\color{orange}u_3 u_1}-47349320565126689937152 {\color{orange}u_1}\nn
&~~~~-821538516578272938693612 {\color{orange}u_2^3}+671461292472253529376858 {\color{orange}u_3^3}\nn
&~~~~+125991080370958589729568 {\color{orange}u_2^2}-2095973906781480200310654 {\color{orange}u_2 u_3^2}\nn
&~~~~+513793121486046706790304 {\color{orange}u_2}+2243470889530529586764928 {\color{orange}u_2^2 u_3}\nn
&~~~~-110384463403936942602144 {\color{orange}u_2 u_3}-493792683630496631192224 {\color{orange}u_3}\big)\co\nn\nn
   \mathcal{F}_{3\to \widehat{23}}(\{u\})&=\frac{4991}{320387178767845798472081158159773726389184}\times\nn
&~~~~\big(3297397614714597368233886798751248713 {\color{orange}u_1^3}\nn
&~~~~-19443287036686359832591721518751736366 {\color{orange}u_2 u_1^2}\nn
&~~~~+10181745500119574531737227437157465027 {\color{orange}u_3 u_1^2}\nn
&~~~~+65393051326809579444635376427293228 {\color{orange}u_1^2}\nn
&~~~~+61926507522720900698260767869192148504 {\color{orange}u_2^2 u_1}\nn
&~~~~+39187119093596304586725478684420390530 {\color{orange}u_3^2 u_1}\nn
&~~~~+46893087661403646892791636313041319968 {\color{orange}u_2 u_1}\nn
&~~~~-90910727085819459452366141566831887444 {\color{orange}u_2 u_3 u_1}\nn
&~~~~-57007506734494284642740432946612237744 {\color{orange}u_3 u_1}\nn
&~~~~-11526804577302466276157420292416003232 {\color{orange}u_1}\nn
&~~~~-157270966175601844250561407992632708889 {\color{orange}u_2^3}\nn
&~~~~+126752157203207979954960713049815650117 {\color{orange}u_3^3}\nn
&~~~~-226128041347371801455874019988009655054 {\color{orange}u_2^2}\nn
&~~~~-399413911378480221235007636494880691357 {\color{orange}u_2 u_3^2}\nn
&~~~~-204172799057499623318436330464241300354 {\color{orange}u_3^2}\nn
&~~~~+37932865128394163521263995655954957696 {\color{orange}u_2}\nn
&~~~~+425455245237593502596410354629123070275 {\color{orange}u_2^2 u_3}\nn
&~~~~+441128347922847494794783646492821079388 {\color{orange}u_2 u_3}\nn
&~~~~-24269275724512593532701830624639304480 {\color{orange}u_3}\big)\ed
\end{align}
The coefficient of $u_3^2$ in $\mathcal{F}_{3\to\widehat{3}}(\{u\})$ is 0 because it contains a factor $(D-6)$. All the above results are consistent with the FIRE6~\cite{Smirnov:2019qkx} and Kira~\cite{Maierhofer:2017gsa}.  In fact we have checked our results for reduction of higher-pole generating functions to master integrals with at most 2 propagators removed also apply to box and pentagon. Due to the similar method and complicated expressions, we do not list them here. We also have checked the results for any sector $n\to n-m$ using box and pentagon.

\section{Discussion}
\label{discussion}
 In this paper, we explore the generating function for higher-pole one-loop integrals, aiming to achieve a comprehensive one-loop reduction generating function. By utilizing reduction techniques in projective space, we derive a partial equation for the higher-pole generating function. From a physics perspective, we gradually eliminate the mass-shifting parameters one by one, thereby transforming the partial differential equations (PDEs) of the generating function into simpler ordinary differential equations (ODEs). The solutions to these ODEs correspond to direct integrals, enabling us to provide a complete solution for the generating function. Consequently, the final outcome can be expressed as a sum of combinations of the functions denoted as ${\cal F}_{n\to \widehat{\boldsymbol{a}}}^{(u_i)}$.
 
 However, this solution fails to exhibit the permutation symmetry explicitly. Interestingly, it is possible to find a simpler integral recursion relation by recognizing the analogy between the PDEs and electromagnetics. By exploiting this analogy, we observe that the obtained result remains unchanged regardless of the integration path chosen. Consequently, the desired permutation symmetry becomes apparent when we select a straight line as the integration path. The expression for ${\cal F}_{n\to \widehat{\emptyset}}$ turns out to be quite straightforward, and the presence of the hypergeometric function arises from the sector denoted as ${\cal F}_{n\to \widehat{b}}$. It is also worth considering the possibility of constructing a generating function for the lower sector using hypergeometric functions, although we do not focus extensively on this issue, as the Taylor series expansion in the integrand proves to be highly useful for calculating a specific higher-pole integral. We anticipate that the method presented in this paper can be applied to higher-loop reductions, leading to higher-order derivations in future investigations. This serves as a reminder for potential avenues of exploration.

\section*{Acknowledgments}
We would like to thank Bo Feng, Chang Hu, Jiyuan Shen, Yaobo Zhang for inspiring discussion and their numerical results of boxes and pentagons for double check. T.Li is supported by NSFC NO. 12175237, Chinese NSF funding under Grant No.11935013, the Fundamental Research Funds for
the Central Universities, and funds from the Chinese Academy of Sciences. Y.Song and L.Zhang are supported by the Fundamental Research Funds for the Central Universities.
\bibliographystyle{JHEP}
\bibliography{reference}
\end{document}